\documentclass[twocolumn, times, tighten]{aastex62}
\usepackage{graphicx, amsmath, amssymb}
\usepackage{multirow} 
\newcommand{\ha}{H\ensuremath{\alpha}}
\newcommand{\hb}{H\ensuremath{\beta}}

\newcommand{\hii}{[H\,{\footnotesize II}]}
\newcommand{\nii}{[N\,{\footnotesize II}]}
\newcommand{\angstrom}{\text{\normalfont\AA}}

\newcommand{\oiii}{[O\,{\footnotesize III}]}

\newcommand{\sii}{[S\,{\footnotesize II}]}

\shorttitle{TDE in NGC\,5092}
\shortauthors{Li et al.}

\begin{document}

\title{Multiwavelength Study of an X-ray Tidal Disruption Event Candidate in NGC 5092}  

\correspondingauthor{Dongyue Li}
\email{dyli@nao.cas.cn}

\correspondingauthor{Weimin Yuan}
\email{wmy@nao.cas.cn}

\author {Dongyue Li}
\affil{Key Laboratory of Space Astronomy and Technology, National Astronomical Observatories, Chinese Academy of Sciences, 20A Datun Road, Chaoyang District, Beijing, 100101, China}
\affil{School of Astronomy and Space Science, University of Chinese Academy of Sciences, 19A Yuquan Road, Beijing, 100049, China}

\author{R.D. Saxton}
\affiliation{Telespazio-Vega UK for ESA, Euporean Space Astronomy Centre, Operations Department, E-28691 Villanueva de la Ca$\tilde{n}$ada, Spain}

\author{Weimin Yuan}
\affil{Key Laboratory of Space Astronomy and Technology, National Astronomical Observatories, Chinese Academy of Sciences, 20A Datun Road, Chaoyang District, Beijing, 100101, China}
\affil{School of Astronomy and Space Science, University of Chinese Academy of Sciences, 19A Yuquan Road, Beijing, 100049, China}

\author{Luming Sun}
\affil{Key Laboratory for Research in Galaxies and Cosmology, Department of Astronomy, The University of Science and Technology of
China, Hefei, Anhui, 230026, China}
\affil{School of Astronomy and Space Sciences, University of Science and Technology of China, Hefei, Anhui, 230026, China}

\author{He-Yang Liu}
\affil{Key Laboratory of Space Astronomy and Technology, National Astronomical Observatories, Chinese Academy of Sciences, 20A Datun Road, Chaoyang District, Beijing, 100101, China}
\affil{School of Astronomy and Space Science, University of Chinese Academy of Sciences, 19A Yuquan Road, Beijing, 100049, China}

\author{Ning Jiang}
\affil{Key laboratory for Research in Galaxies and Cosmology, Department of Astronomy, The University of Science and Technology of
China, Hefei, Anhui, 230026, China}
\affil{School of Astronomy and Space Sciences, University of Science and Technology of China, Hefei, Anhui, 230026, China}

\author{Huaqing Cheng}
\affil{Key Laboratory of Space Astronomy and Technology, National Astronomical Observatories, Chinese Academy of Sciences, 20A Datun Road, Chaoyang District, Beijing, 100101, China}
\affil{School of Astronomy and Space Science, University of Chinese Academy of Sciences, 19A Yuquan Road, Beijing, 100049, China}

\author{Hongyan Zhou}
\affil{Antarctic Astronomy Research Division, Key Laboratory for Polar Science of the State Oceanic Administration, Polar Research Institute of China, 451 Jinqiao Road, Shanghai, 20136, China}
\affil{School of Astronomy and Space Sciences, University of Science and Technology of China, Hefei, Anhui, 230026, China}
\affil{Key laboratory for Research in Galaxies and Cosmology, Department of Astronomy, The University of Science and Technology of
China, Hefei, Anhui, 230026, China}

\author{S. Komossa}
\affiliation{Max Planck Institut f\"ur Radioastronomie, Auf dem H\"ugel 69, D-53121 Bonn, Germany}
\affil{Key Laboratory of Space Science Astronomy and Technology, National Astronomical Observatories, Chinese Academy of Sciences, 20A Datun Road, Chaoyang District, Beijing, 100101, China}

\author{Chichuan Jin}
\affil{Key Laboratory of Space Astronomy and Technology, National Astronomical Observatories, Chinese Academy of Sciences, 20A Datun Road, Chaoyang District, Beijing, 100101, China}
\affil{School of Astronomy and Space Science, University of Chinese Academy of Sciences, 19A Yuquan Road, Beijing, 100049, China}

\begin{abstract}
We present multiwavelength studies of a transient X-ray source, XMMSL1\,J131952.3+225958,
associated with the galaxy NGC\,5092 at $z=0.023$ detected in the {\it XMM-Newton} SLew survey (XMMSL). 
The source brightened in the 0.2--2\,keV band by a factor of $>20$ in 2005 as compared with previous flux
limits and then faded by a factor of $>200$ as observed with {\it XMM-Newton} in 2013 and with {\it Swift} in 2018. 
At the flaring state, the X-ray spectrum can be modeled with a blackbody at a temperature of $\sim$\,60\,eV and
an overall luminosity of $\sim$\,$1.5 \times 10^{43}$\,erg\,s$^{-1}$. A UV flare and optical flare were also
detected with the {\it Galaxy Evolution Explorer} and the Sloan Digital Sky Survey, respectively, 
within several months of the X-ray flare, whose nonstellar UV--optical spectrum can be described with
a blackbody at a temperature of $\sim$\,$(1-2) \times 10^4$\,K and a luminosity of $\sim$\,$(2-6) \times 10^{43}$\,erg\,s$^{-1}$.
 Interestingly, mid-infrared monitoring observations of NGC\,5092 with the {\it Wide-field Infrared Survey Explorer} 5--13\,yr
later show a continuous flux decline. These dramatic variability properties, from the X-ray through UV and optical to infrared,
appear to be orderly, suggestive of a stellar tidal disruption event (TDE) by a massive black hole,
confirming the postulation by \citet{2013ApJ...774...63K}. This TDE candidate belongs to a rare sample
with contemporaneous bright emission detected in the X-ray, UV, and optical, which are later echoed by
dust-reprocessed light in the mid-infrared. The black hole has a mass of $\sim$\,$5 \times 10^{7} \rm M_{\odot}$,
residing in a galaxy that is dominated by a middle-aged stellar population of 2.5\,Gyr.
\end{abstract}


\section{Introduction} 
A tidal disruption event (TDE) occurs when a star passes so close to a supermassive black hole (SMBH) 
at the center of a galaxy that the tidal force of the SMBH is able to overcome the self-gravity of
the star and disrupt it. During this process, roughly half of the stellar material is accreted
onto the SMBH, resulting in a flare of radiation peaking in soft X-ray and UV band,
fading away on time scales of months to years (\citealp{1988Natur.333..523R}). 
These events are a unique probe of quiescent SMBHs at the center of normal galaxies,
which are otherwise difficult to detect and may contribute to the growth and evolution of SMBHs.
Meanwhile, TDEs are an ideal laboratory for studying the accretion around SMBHs,
as we may witness the whole process where the accretion rate varies from low to super-Eddington,
which normally takes millennia or more in active galactic nuclei (AGN; e.g., \citealp{2015JHEAp...7..148K}).
In TDEs, since the debris of the stars are accreted to the BHs from a close distance, the compact TDE disks will emit mostly X-rays.
The first TDE was detected in soft X-ray in the \hii-like nucleus of NGC 5905 (\citealp{1996A&A...309L..35B}; \citealp{1999A&A...343..775K}).
The X-ray luminosity showed a variability amplitude larger than a factor of 150 within 2\,yr, 
with an extremely soft X-ray spectrum in the high state (photon index $\Gamma$\,$\sim$\,4).
All these properties are consistent with the expectations of the TDE model. So far, 
only about two dozen TDEs have been discovered by the detection of their X-ray flares
(e.g., \citealp{1999A&A...349L..45K, 2008A&A...489..543E, 2011ApJ...738...52L, 2012A&A...541A.106S, 
2017A&A...598A..29S, 2019A&A...630A..98S}, refer to \citealp{2015JHEAp...7..148K} for a recent review). 
The observed X-ray light curves of TDEs typically follow a power-law decline 
approximately by $t^{-5/3}$ over timescales of months to years,  
roughly consistent with the theoretical prediction (e.g., \citealp{1988Natur.333..523R}). 
Their X-ray spectra are mostly very soft, with power-law photon index $\Gamma$\,$\sim$\,4--5, 
and can be described with blackbody emission with a temperature around 40--100\,eV 
(corresponding to $\sim$\,$10^5-10^6$\,K; \citealp{2002RvMA...15...27K}). 
There is also a class of transients that shows luminous and energetic flaring 
in hard X-ray and $\gamma$-ray bands accompanied by radio emission. Here, the rapid variability, 
huge X-ray peak luminosity, blazar-like spectral energy distribution (SED), and optically inactive host galaxy
can all be explained by the launch of a relativistic jet following the tidal disruption of a star. 
TDEs with jets could provide new insight into the formation and launch of radio jets 
(e.g. \citealp{2011Sci...333..203B}; \citealp{2011Natur.476..421B}; \citealp{2011Sci...333..199L, 2011Natur.476..425Z}; \citealp{2012ApJ...753...77C}; \citealp{2015MNRAS.452.4297B}). 

In recent years, thanks to the development of high-cadence optical and UV surveys, 
a number of TDEs have been found as optical and UV transients on time scales of months to years 
(\citealp{2006ApJ...653L..25G, 2008ApJ...676..944G, 2012Natur.485..217G}; \citealp{2014MNRAS.445.3263H}; 
\citealp{2017ApJ...842...29H}; \citealp{2019ApJ...878...82V, 2019ApJ...883..111H}; see \citealp{2018ApJ...852...72V} for a recent compilation).  
The optical and UV SED can normally be fitted with blackbody radiation at temperatures $\sim$\,(1--3)\,$\times$\,10$^4$\,K (\citealp{2018ApJ...852...72V}).
The origin of the optical and UV emission in TDEs is unclear yet, as the peak temperature is much lower than that 
expected from the canonical TDE models (e.g., 10$^5$--10$^6$\,K). Current models for the optical emission mainly include
(1) thermal emission from a disk (\citealp{2011MNRAS.410..359L}); (2) emission during the formation of the accretion disk, 
in which the emission is powered by shocks from intersecting stellar debris streams (\citealp{2015ApJ...806..164P}); 
and (3) reprocessing of the hotter disk emission by the outer envelope, where the envelope is formed by outflow during 
the accretion process (\citealp{2014ApJ...783...23G}; \citealp{2016MNRAS.461..948M}) or stellar debris (\citealp{1997ApJ...489..573L}).
Only a few TDEs have contemporaneous X-ray and optical detections, e.g., 
ASASSN-14li (\citealp{2015Natur.526..542M}; \citealp{2016ApJ...818L..32C}; \citealp{2016MNRAS.455.2918H}),
 XMMSL1\,J074008.2-853927 (\citealp{2017A&A...598A..29S}), AT2018fyk (\citealp{2019MNRAS.488.4816W}), 
AT2018zr (\citealp{2019ApJ...878...82V}), and ASASSN-19bt (\citealp{2019ApJ...883..111H}). A unified, 
viewing-angle-dependent model has been proposed by \citet{2018ApJ...859L..20D} for interpreting the various 
classes of observed TDEs. In this model, the X-ray emission, originating from the inner disk, would be obscured
by the optically thick super-Eddington disk wind when viewed close to the disk plane, and optical TDEs are
produced by the reprocessing of the X-ray photons. When viewing high above the disk, where the gas 
is optically thin to absorption, one would expect to see both the X-ray and optical/UV emission. 
The detection of strong Bowen fluorescence lines in the optical spectrum of several TDEs with
X-ray nondetections supports this scenario (\citealp{2019ApJ...887..218L}).

TDEs in gas-rich environments may illuminate circum-nuclear material, and 
the reprocessed emission lines provide us valuable information for the study of the cores of  
quiescent galaxies (e.g., \citealp{2008ApJ...678L..13K}; \citealp{2011ApJ...740...85W, 2012ApJ...749..115W}).
 If the TDE takes place in a dusty environment, most of the UV/optical photons 
will be absorbed by dust grains in the vicinity of the BH and reradiated in infrared (IR),
which is the IR echo of TDE (\citealp{2009ApJ...701..105K}; \citealp{2016ApJ...832..188D, 2017ApJ...841L...8D}; 
\citealp{2016ApJ...828L..14J, 2017ApJ...850...63J}; \citealp{2016MNRAS.458..575L}; \citealp{2016ApJ...829...19V, 2018MNRAS.477.2943W}).
The TDE research is currently hampered by the small number of known events. Many questions remain, 
including the event rate and luminosity function of TDEs, origin of the diversity of two classes of TDEs, 
how the jets form, the relation of the TDE incidence to the host galaxy, and the star forming history. 
To answer these questions, a large number of TDEs are needed, especially those showing X-ray flares.

The {\it XMM-Newton} SLew survey (XMMSL; \citealp{2008A&A...480..611S}) contains data taken with the EPIC-pn camera during slews
from one pointed observation to another. In the second XMMSL catalog (XMMSL2),\footnote{\url{https://www.cosmos.esa.int/web/xmm-newton/xmmsl2-ug}} 84\% of the 
sky has been covered with a flux limit of 6\,$\times$\,10$^{-13}$\, erg\,s$^{-1}$\,cm$^{-2}$ 
and a mean exposure time of 6\,s in the energy band of 0.2--2\,keV. 
Comparison of source fluxes in the XMMSL survey with those measured 
in the previous ROSAT All-Sky Survey (RASS; \citealp{1999A&A...349..389V}) performed in the 1990s has allowed 
the detection of five sources that are suggested to be TDEs: XMMSL1\,J1115+1806 (\citealp{2008A&A...489..543E}),
XMMSL1\,J1323+4827 (\citealp{2008A&A...489..543E}), XMMSL1\,J1201+3003 (\citealp{2012A&A...541A.106S}), 
XMMSL1\,J0740-8539 (\citealp{2017A&A...598A..29S}), and XMMSL2 J1446+6857 (\citealp{2019A&A...630A..98S}).
All five of these TDEs show soft X-ray flux variations greater than a factor of\,50 in optically inactive galaxies. 
For three (XMMSL1\,J1115+1806, XMMSL1\,J1323+4827, XMMSL\,J1201+3003) of these five TDEs, the outburst X-ray spectra are very soft, 
with photon index $\Gamma$\,$\sim$\,3--4 if fitted with a single power-law. TDE XMMSL1\,J0740-8539 shows both a thermal and
a non-thermal component with a spectrum that may be modeled with thermal emission in the UV band, a power law with $\Gamma$\,$\sim$\,2 
dominating in the X-ray band above 2\,keV, and a soft X-ray excess with an effective temperature of $\sim$\,86\,eV. 
XMMSL2\,J1446$+$6857 has flat X-ray flux for $\sim$\,100 days and then falls by a factor of 100 over the following 500 days. 
Throughout the evolution of the event, the X-ray spectrum can be modeled with a power law of $\Gamma$\,$\sim$\,2.6 
and may be solely due to Compton upscattering of thermal photons from the disk.

In this paper, we present the multiwavelength studies of a TDE candidate, XMMSL1\,J131952.3+225958 (hereafter XMMSL1 J1319+2259), 
associated with the galaxy NGC\,5092 at a redshift of $z = 0.023$. The X-ray source was first detected in the XMMSL survey in 2005. 
It was noted as a possible TDE candidate in a sample study of X-ray transients selected from the XMMSL by \citet{2013ApJ...774...63K}, 
based on variability by a factor of $>$10 compared to the previous RASS upper limit, though there was a lack of further compelling evidence. 
Its X-ray flux was found to be $\sim$\,240 times fainter in a pointed observation with {\it XMM-Newton} in 2013. 
The Sloan Digital Sky Survey (SDSS; \citealp{2000AJ....120.1579Y}) and {\it Galaxy Evolution Explorer} ({\it GALEX}; 
\citealp{2005ApJ...619L...7M}, \citealp{2005ApJ...619L...1M}) observations in 2005 show that there was
also an optical and UV flare in the nucleus of NGC 5092. This galaxy showed a continuous decline
of the {\it Wide-field Infrared Survey Explorer} ({\it WISE}; \citealp{2010AJ....140.1868W}) infrared flux from 2010 to 2018, 
suggesting an infrared echo following the optical/UV/X-ray flare.

In Section \ref{x-ray data}, we describe the X-ray observations, data reduction, and analysis of XMMSL1 J1319+2259. 
In Section \ref{data reduction}, we compile and study the multiwavelength observations, from IR to UV, 
of the nuclear and host galaxy of this source, and an optical spectral analysis is performed in Section \ref{sec:optical spectra}. 
The results are summarized and discussed in Section \ref{discussion}. 
Details of the data reduction and analysis are described in Appendix. 

Throughout this paper, we adopt a flat cosmology with $H_0= 70$\,km\,s$^{-1}$\,Mpc$^{-1}$, $\Omega_m=0.3$, and $\Omega_A = 0.7$.
The redshift of NGC\,5092 ($z=0.023$) corresponds to a luminosity distance of 100.2 Mpc.

\section{X-Ray data analysis}
\label{x-ray data}

\subsection{The X-Ray Source XMMSL1\,J1319+2259 and X-Ray Observations}

\begin{figure*}
 \centering
 \includegraphics[width=0.6\textwidth]{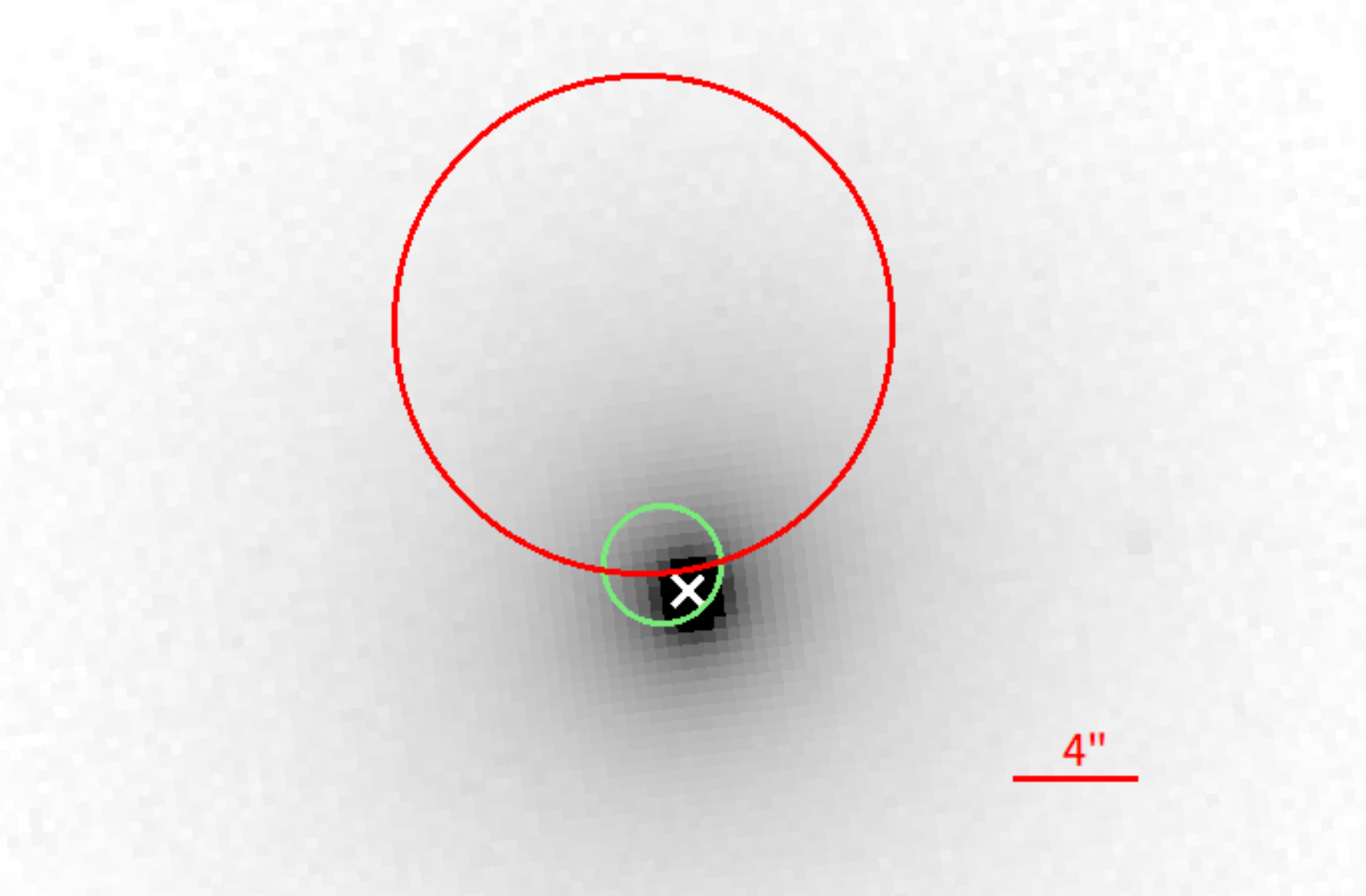}
 \caption{An $r$-filter image of the galaxy NGC~5092 from the SDSS, taken on 2005 April 2, with the center of the galaxy
marked with a white cross. Also shown is the {\it XMM-Newton} slew error circle (red; 8\arcsec at the 68\% confidence level) and {\it XMM-Newton} pointed error
circle (green; 1.9\arcsec at the 68\% confidence level).}
 \label{fig:galaxy_2}
\end{figure*}

The source XMMSL1\,J1319+2259 was first detected 
in the soft X-ray band (0.2--2\,keV) on 2005 July 15  during 
an XMM slew observation, with an exposure time of 4.6~s.
The source position, taken from the XMMSL2 catalog, is at R.A.=$13^{\rm h}19^{\rm m}52^{\rm s}.2$, decl.=$22^\circ 59^{'}58^{''}.2$ (J2000), 
with a position error of 8\arcsec in radius at the 68\% confidence level (\citealp{2008A&A...480..611S}). 
Within this error circle, there is only one cataloged bright source (see Figure \ref{fig:galaxy_2}), 
the galaxy NGC\,5092, at R.A.=$13^{\rm h}19^{\rm m}51^{\rm s}.5$, decl.=$22^\circ 59^{'}59^{''}.6$ (\citealp{2018A&A...616A...1G}), 
with a redshift of 0.023.  By analyzing the XMMSL data, we find that the source is detected only in the 0.2--2\,keV soft X-ray band 
with a count rate of $3.0\pm 0.8$\,counts\,s$^{-1}$, and has a $2\sigma$ upper limit of 1.0\,counts\,s$^{-1}$ in the 2--10\,keV band, 
indicative of a soft X-ray spectrum. There was another slew coverage of NGC~5092 in 2002 with an exposure time of 10\,s, 
setting an upper limit count rate of 0.4\,counts\,s$^{-1}$ ($<5.2\,\times$\,10$^{-13}$\,erg\,s$^{-1}$\,cm$^{-2}$; 
see Section \ref{sec.x-ray spectra}) in the 0.2--2\,keV band. The galaxy NGC 5092 was not detected in the RASS in 1990 July, 
which yields an upper limit of 0.04\,counts\,s$^{-1}$ in the 0.2--2\,keV band, or $3.5\,\times\,10^{-13}$\,erg\,s$^{-1}$\,cm$^{-2}$
in flux (Section \ref{sec.x-ray spectra}). This sets a stringent upper limit on the X-ray luminosity of 4.2\,$\times\,10^{41}$\,erg\,s$^{-1}$,
indicating that NGC 5092 does not host a strong AGN. 
An {\it XMM-Newton} pointed observation with an exposure of $\sim$\,8600\,s was carried out on 2013 December 21 (PI: Richard Mushotzky; ObsID=0721830301).
We perform data reduction following the standard procedures with the software {\it XMM-Newton} Science Analysis System 
(\textsc{sas} version 16.0.0; \citealp{2004ASPC..314..759G}), and the details are described in Appendix \ref{sec:app.x-ray}.
The refined source position given in the 3XMM-DR8 (\citealp{2016A&A...590A...1R}) is R.A.=$13^{\rm h}19^{\rm m}51^{\rm s}.6$, decl.=$22^\circ 59^{'}58^{''}.8$,
 with a source position error of 1.9\arcsec at the 68\% confidence level (shown as  green circle in Figure \ref{fig:galaxy_2}).
Detailed observational information of the three detectors (pn, MOS1, and MOS2), including the good exposure time and count rates, 
is summarized in Table \ref{X-ray observation}.

There was also an observation performed by the X-ray Telescope (XRT; \citealp{2005SSRv..120..165B}) on board 
the Neil-Gehrels {\it Swift} Observatory (\citealp{2004ApJ...611.1005G}) in 2018 August with an exposure time 
of 576\,s, setting a $2\sigma$ upper limit of 0.005 counts s$^{-1}$, or $1.6 \times 10^{-13}$\,erg\,s$^{-1}$\,cm$^{-2}$ 
assuming a power-law spectral model with $\Gamma=2.1$ modified by Galactic absorption (see Section \ref{sec.x-ray spectra}).
Table \ref{X-ray observation} summarizes the information of all the available X-ray observations of this source and its detection. 
Among the five observations, the source was detected in two, namely, the XMMSL in 2005 and the pointed XMM observation in 2013.

\begin{deluxetable*}{ccccccc}
\tablecolumns{7}
 \tablecaption{Log of all the XMMSL1 J1319+2259 X-Ray observations}
 \tablehead{\colhead{Obs. Date} & \colhead{ObsID} & \colhead{Mission/Ins.} & \colhead{Exposure} & \colhead{Energy Band} & \colhead{Count Rate}& \colhead{Unabs. Flux} \\
 & & & (s) & (keV) & (counts s$^{-1}$) & (10$^{-13}$\,erg\,s$^{-1}$ cm$^{-2}$) }
\startdata
\hline
1990 Jul 11& & ROSAT& 337& 0.2--2& $<0.04$ & $<3.5$  \\
\hline
2002 Jul 06& 9047100005 & XMM/pn& 10& 0.2--2&  $<0.4$ & $<5.2 $\\
\hline
2005 Jul 15& 9102500005 & XMM/pn& 5& 0.2--2& $3.0\pm 0.8$ & $70.9^{+29.8}_{-22.4} $ \\
\hline
2013 Dec 21& 0721830301 & XMM/pn& 1467& 0.2--2& $(8.9\pm 2.6)\times 10^{-3}$ & $0.3^{+0.1}_{-0.1}$\\
$''$&$''$&$''$&$''$&2--8&$(2.0\pm 1.8)\times 10^{-3}$& $0.2^{+0.03}_{-0.04}$ \\
 $''$&$''$& XMM/MOS1&7182&0.2--2& $(2.6\pm 0.7)\times 10^{-3}$ & $0.3^{+0.1}_{-0.1}$\\
$''$&$''$&$''$&$''$&2--8&$(0.4\pm 0.2)\times 10^{-3}$& $0.2^{+0.03}_{-0.04}$\\
$''$&$''$& XMM/MOS2&7157&0.2--2& $(0.9\pm 0.5)\times 10^{-3}$ & $0.3^{+0.1}_{-0.1}$\\
$''$&$''$&$''$&$''$&2--8&($0.7\pm 0.4)\times 10^{-3}$ & $0.2^{+0.03}_{-0.04}$ \\
\hline
2018 Aug 24& 03106172001& {\it Swift}/XRT & 576& 0.2--2& $<0.005$& $<1.40$ \\
\hline
\enddata
\tablecomments {The unabsorbed fluxes with $1\sigma$ error and $2\sigma$ upper limits are estimated using \textsc{xspec}
as described in Section \ref{sec.x-ray spectra}.}
\label{X-ray observation}
\end{deluxetable*}

\subsubsection{X-Ray Spectral Analysis}
\label{sec.x-ray spectra}
Here we perform spectral analysis for the XMMSL and XMM pointed observation using \textsc{xspec} (version 12.9.1; \citealp{1996ASPC..101...17A}).
We extract photons detected in the range of 0.2--10\,keV, where the data are best calibrated and are in the most 
sensitive range for {\it XMM-Newton} EPIC cameras. For the XMM pointed observation, the MOS1 and MOS2 spectra are combined, 
taking into account the responses, background, and ancillary files. The combined MOS and the pn spectrum are jointly 
fitted in all the analyses below. Given the small number of photons of the detections, the combined MOS spectrum is 
binned with a minimum of three counts per bin, and the pn spectra in the XMMSL and XMM pointed 
observation are binned to have only one photon per group in 0.2--10\,keV,
and the {\it C}-statistic is used for the fitting. 
The Galactic equivalent $N_{\rm H}$ column density toward NGC~5092 is $1.22 \times 10^{20}$\,cm$^{-2}$ 
(\citealp{2013MNRAS.431..394W}), which is always taken into account and modeled by \texttt{tbabs} with the
element abundances set to those in \citet{2000ApJ...542..914W}.  
We find that in all the spectral fits below, no additional absorption, neutral or ionized, is required. 
The fitting results are summarized in Table \ref{X-ray  spectra parameters}. Note that the quoted uncertainties 
of the fitted parameters are at the 90\% confidence range.

\begin{deluxetable*}{ccccc}
 \tablecolumns{5} 
 \tablecaption{Spectral fitting to the XMMSL and XMM pointed observation of XMMSL1 J1319+2259.}
\tablehead{\colhead{Obs}  & \colhead{Model}&\colhead{ $\Gamma$}& \colhead{$kT (\rm eV)$} &\colhead{ $C$ (dof)} }
\startdata
 \hline
 pointed& \texttt{tbabs*(zpo)}&  $2.1^{+0.5}_{-0.5}$ & $(-)$  & 17.5(33) \\
 \hline
 slew&\texttt{tbabs*(zpo)}& $4.4^{+2.0}_{-1.7}$ & $(-)$ & 8.9(11) \\
 $''$   &\texttt{tbabs*(zbb)} & $(-) $ &$64.2^{+31.2}_{-22.0}$ & 9.2(11)\\
 $''$    &\texttt{tbabs*(zpo+zbb)}&2.1(fixed) & $64.0^{+31.2}_{-22.1}$ & 9.2(11)\\
 \hline
\enddata
\tablecomments{ All fits include absorption by Galactic column (model \texttt{tbabs}, $N_{\rm H}$=1.22$\times 10^{20}$ atoms cm$^{-2}$,
 abundances from \citealp{2000ApJ...542..914W}). For the XMMSL spectrum, while fitted with model \texttt{tbabs}*(\texttt{zpo}+\texttt{zbb}), parameters
of the power law model are fixed at values from XMM pointed spectrum fitting. Errors are at 90\% confidence level.} 
\label{X-ray spectra parameters}
\end{deluxetable*}


For the XMM pointed observation, a power law modified by  Galactic absorption, \texttt{tbabs$\times$zpowerlaw}, 
provides an acceptable fit to the pn+MOS spectrum with a photon index $\Gamma=2.1^{+0.5}_{-0.5}$. 
The corresponding unabsorbed (corrected for Galactic absorption) 0.2--2\,keV flux is $0.3^{+0.1}_{-0.1} \times 10^{-13}$\,erg\,s$^{-1}$\,cm$^{-2}$, 
with a rest-frame luminosity of $3.6^{+1.2}_{-1.2} \times 10^{40}$\,erg\,s$^{-1}$. 
The Eddington ratio $L_{0.2-2\,{\rm keV}}/L_{\rm Edd}$ is estimated to be $\sim$\,$5.7 \times 10^{-6}$ 
($M_{\rm BH}=5.0^{+5.6}_{-2.9} \times 10^{7} {\rm M}_{\odot}$, please refer to Section \ref{sec:optical spectra} 
for the BH mass estimation). We note that this value should be an upper limit, as the emission from the host galaxy 
may significantly contribute to the detected flux at such a low luminosity level.

\begin{figure*}
 \centering
 \includegraphics[width=0.5\textwidth]{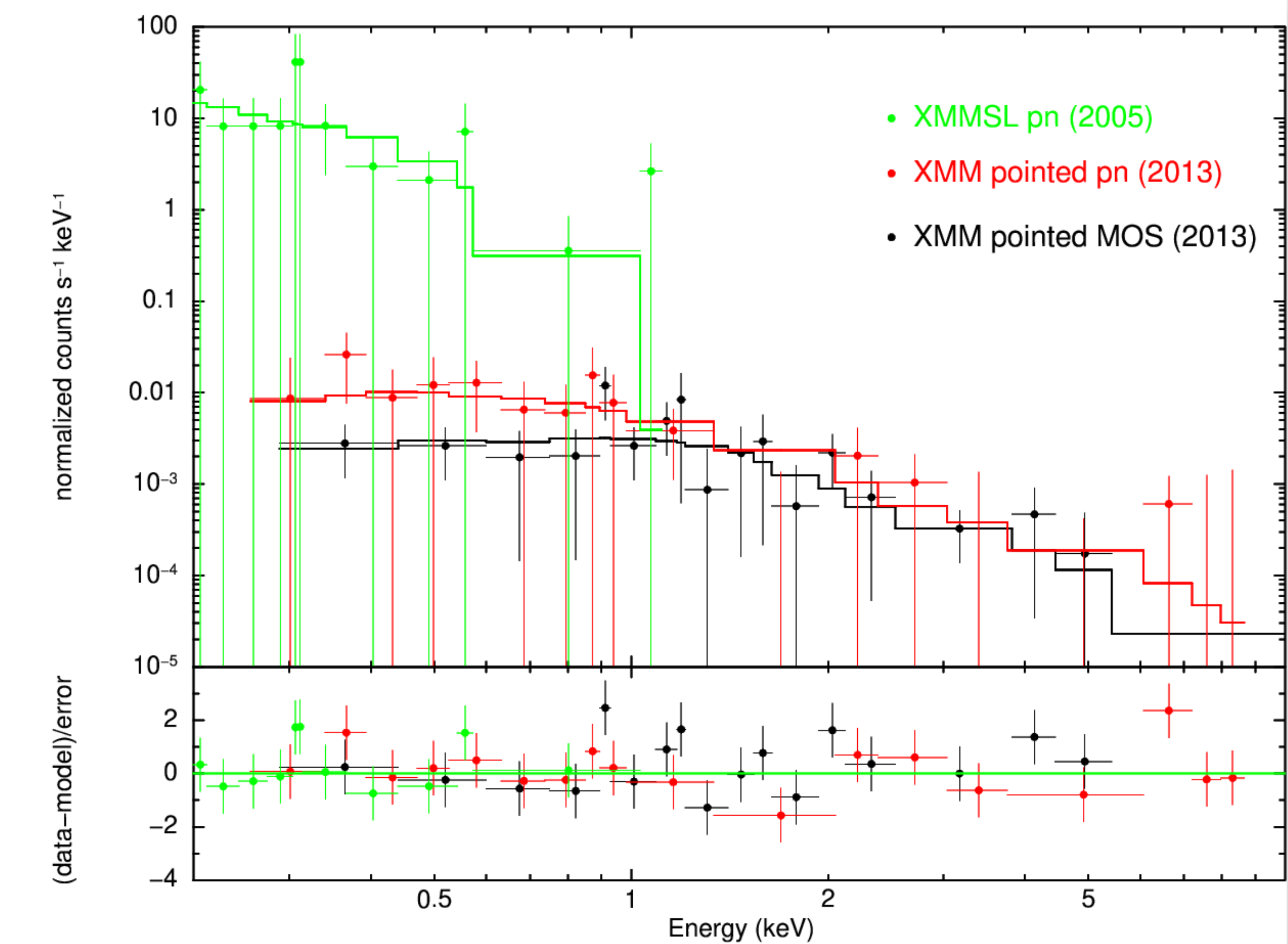}
 \caption{The 0.2-10 keV pn (red) and combined MOS (black) spectra obtained from XMM pointed (red) observations in 2013, 
and the pn spectrum (green) from the XMMSL in 2005. The XMM pointed pn and combined MOS spectra are fitted simultaneously 
with an absorbed power-law model and the XMMSL spectrum is fitted with a blackbody model, 
as described in Section \ref{sec.x-ray spectra}.}
 \label{fig:x-ray spectra}
\end{figure*}

For the pn spectrum taken in the XMMSL observation, which is very soft, we first try two simple models of the source 
emission: a power law (\texttt{zpowerlaw}) and a blackbody (\texttt{zblackbody}).
The blackbody model provides a good fit to the data, yielding a temperature of $kT = 64.2^{+31.2}_{-22.0}$\,eV, 
which is consistent with measurements from previous TDEs (e.g., \citealp{2002RvMA...15...27K}). In the rest-frame $0.2-2$~keV band, the unabsorbed 
flux is $70.9^{+29.8}_{-22.4} \times 10^{-13}$\,erg\,s$^{-1}$ cm$^{-2}$, corresponding to a luminosity of $8.6^{+3.6}_{-2.7} \times 10^{42}$\,erg\,s$^{-1}$, 
and thus an Eddington ratio $L_{0.2-2~{\rm keV}}/L_{{\rm Edd}} \sim 1.4 \times 10^{-3}$. 
The overall luminosity of the blackbody emission is $\sim$\,$1.5 \times 10^{43}$\,erg\,s$^{-1}$  
($L_{{\rm 0.01-100~{\rm keV}}}/L_{{\rm Edd}}\sim\,2.4 \times 10^{-3}$). A power-law model also gives 
an equally good fit with $\Gamma=4.4^{+2.0}_{-1.7}$. We also try to fit the spectrum with a blackbody plus a power-law model, 
but find that the addition of a power law component does not improve the fit.
Figure \ref{fig:x-ray spectra} shows the two X-ray spectra and model fits. 
The spectrum became harder from 2005 to 2013, and the flux dropped by a factor of $\sim$\,240.

For the nondetections in the remaining observations, the flux limits are estimated (see Table \ref{X-ray observation}) 
by assuming the source spectra to be the same as those found in the XMM pointed observation, since 
the source appeared to be at a low state in all these observations.

\subsubsection{X-Ray light curve}
\begin{figure*}
 \centering
 \includegraphics[width=0.6\textwidth]{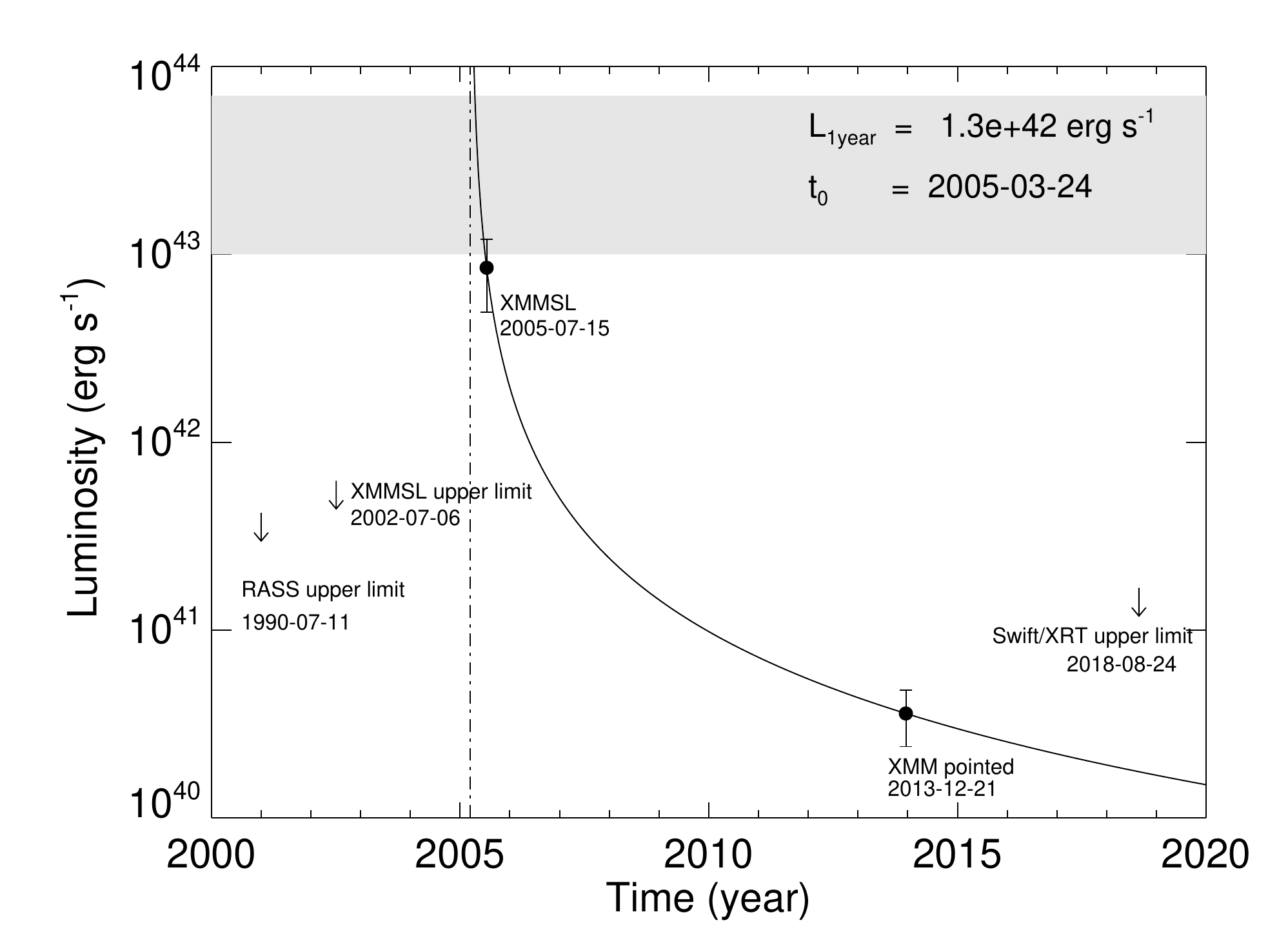}
 \caption{X-ray light curve of XMMSL1 J1319+2259  as the luminosities in 0.2--2\,keV (circles) and upper limits (arrows). 
A t$^{-5/3}$ decline law is fitted to the detected data points: $L(t)=L_{\rm 1year}\times [\frac{ t-t_0}{\rm 1yr}]^{-5/3} \mathrm{erg \ s}^{-1}$, 
where $t_0$ is the tidal disruption time and $L_{\rm 1year}$ is the luminosity 1 yr after the disruption. 
The best-fit parameters are shown in the figure, and the solid curve is the curve predicted for the decline stage.
The dashed line shows the disruption time ($t_0$) obtained from the fitting. The region in light gray shows the 
range of the peak luminosity assuming that the flare began to decrease 
at an epoch from 1 to 3 months after the disruption.}
 \label{fig:x-ray light curve}
\end{figure*}

The obtained unabsorbed 0.2--2\,keV fluxes and upper limits are plotted in Figure \ref{fig:x-ray light curve}.
The flux increased by a factor of more than 20 in 2005 compared to the RASS observation and then dropped by a factor of $\sim$\,240 from 2005 to 2013
as observed in the XMM pointed observation, remaining in a low state till the {\it Swift} observation in 2018. 
Though the data points are scarce, such a large-amplitude X-ray flare resembles those of the known X-ray-selected TDEs (see \citealp{2015JHEAp...7..148K} for review).
In the process of tidal disruption, the X-ray flare is ignited when the most bound material returns to the pericenter after completing
one orbit. Assuming that the material is uniformly distributed in the disrupted stellar debris,  as 
demonstrated by numerical simulations, the mass fallback rate after one post-disruption orbit is
\begin{equation}
\ \dot M\propto\left[\frac{t-t_{\rm 0}}{1 \rm yr.}\right]^{-5/3},
\end{equation}
where $t_{0}$ is the time at which the disruption process occurs (\citealp{1989ApJ...346L..13E}; \citealp{2000ApJ...545..772A}). 
The X-ray radiation power is assumed to evolve following this $t^{-5/3}$ law.
Assuming that this is a TDE and the debris evolution follows the above fallback model, the luminosity declines following the  
canonical $t^{-5/3}$ law
\begin{equation}
\ L(t)=L_{\rm 1year} \times \left [\frac{ t-t_0}{\rm 1yr}\right]^{-5/3} \mathrm{erg \ s}^{-1},
\label{eq:lc}
\end{equation}
where $L_{\rm 1year}$ is the luminosity 1 yr after the disruption occurring at $t_0$. 
The two parameters, $L_{\rm 1year}$ and $t_0$ can be determined by using the two detected data points (luminosities), 
giving $L_{\rm 1year}=(1.3\pm 0.5)\times 10^{42}$\,erg\,s$^{-1}$ and $t_0=2005.2 \pm 0.1$, i.e. the disruption time is 
somewhere around 2005 February to April. Assuming that the flux begins to decrease at an epoch from about 1 to 3 months 
after the disruption,\footnote{This has been observed to be the case for TDEs detected in the optical/UV band (\citealp{2013MNRAS.435.1809S}, \citealp{2019ApJ...878...82V}), 
and we assume that the X-ray rising and peaking timescales are comparable to those of the optical/UV flares.} the derived peak luminosity in $0.2-2$\,keV
is in the range of $\sim$\,(1--7)\,$\times 10^{43}$\,erg\,s$^{-1}$.
We note that the X-ray properties of this variable source, the amplitude of the flare, the very soft spectral shape, 
and the peak luminosity are typical of the TDEs known so far.

\section{Optical, UV, and infrared data analysis} \label{data reduction}

The galaxy NGC~5092 was covered in several surveys, e.g., the All-sky Imaging Survey (AIS) conducted by {\it GALEX} in UV, 
the SDSS in optical, the Two-Micron All Sky Survey (2MASS; \citealp{2006AJ....131.1163S}) in near-infrared (NIR),
and the all-sky survey performed by \texttt{WISE} in mid-infrared (MIR). The galaxy was also observed by the 
optical/UV instruments onboard the {\it XMM-Newton} and {\it Swift} during the X-ray pointed observations.
An optical spectrum was taken in the SDSS in 2008. Information of the multiwavelength observations, 
from far-UV (FUV) to IR, and the photometry results are summarized in Table \ref{tab:photometry}.
In this section, we describe the data reduction and analysis, and some of the technique details can be 
found in Appendix \ref{sec:appendix} for each of the instruments.

\begin{deluxetable*}{cccccc}
 \tablecolumns{6} 
\tablecaption{Log of UV, optical and IR observations and photometric results of NGC~5092}
\tablehead{\colhead{Obs. Date}& \colhead{Survey or Mission} & \colhead{Filter} & \colhead{Exposure (s)}& \colhead{Magnitude}& \colhead{Aperture (arcsec)}}
\startdata
\hline
2007 May 6 & {\it GALEX} & FUV &108& $20.16\pm 1.14$ & 20\\
2005 May 28 & {\it GALEX} & NUV &108& $15.51 \pm 0.01$ & 20\\
2005 Jun 11 & {\it GALEX} & NUV &90& $15.45 \pm 0.02$ & 20\\
2007 May 6 & {\it GALEX} & NUV &106& $18.66 \pm 0.08$ & 20\\
2018 Aug 24 & {\it Swift} (UVOT) & UW1&288&$17.55 \pm 0.09$  & 20\\
2018 Aug 24 & {\it Swift} (UVOT) & $U$ &274& $15.78 \pm 0.02$  & 30\\
2013 Dec 21 & XMM (OM) & $U$ &5000& $15.67 \pm 0.01$ & 20\\
2005 Mar 10 & SDSS DR7 & $u$ &54& $15.07 \pm 0.01$  & 30\\
2005 Mar 10 & SDSS DR7 & $g$&54& $13.71 \pm 0.01$  &35\\
2005 Mar 10 & SDSS DR7 & $r$ &54& $12.96 \pm  0.02$ & 35\\
2005 Mar 10& SDSS DR7 & $i$ &54& $12.59 \pm 0.02$  & 35\\
2005 Mar 10 & SDSS DR7 &$z$ &54& $12.37 \pm 0.01$ & 35\\
2005 Apr 2 & SDSS DR7 & $u$&54& $14.81 \pm 0.01$ &  30\\
2005 Apr 2 & SDSS DR7 &$g$ &54& $13.54 \pm 0.01$ & 35\\
2005 Apr 2 & SDSS DR7&$r$ &54& $12.98 \pm 0.02 $& 35\\
2005 Apr 2 & SDSS DR7&$i$ &54& $12.59 \pm 0.02$  &35\\
2005 Apr 2 & SDSS DR7&$z$ &54& $12.34 \pm 0.01$  & 35\\
1999 Jun 8 & 2MASS & $J$ &355& $10.99 \pm 0.08$  &  30\\
1999 Jun 8 & 2MASS & $H$ &355& $10.25 \pm 0.09$  & 30\\
1999 Jun 8 & 2MASS & $K_{\rm s}$ &355& $9.88 \pm 0.15$ & 30\\
\hline
\enddata
\tablecomments{ This table gives the aperture photometric magnitude of the whole galaxy in multiple bands within 
the apertures of radii listed in the last column. 
All the magnitudes are in the AB magnitude system.}
\label{tab:photometry}
\end{deluxetable*}

\subsection{Optical imaging and photometric data analysis}
\subsubsection{SDSS Imaging Data}
\label{sdss image}
\begin{figure*}
\centering
\includegraphics[width=0.8\textwidth]{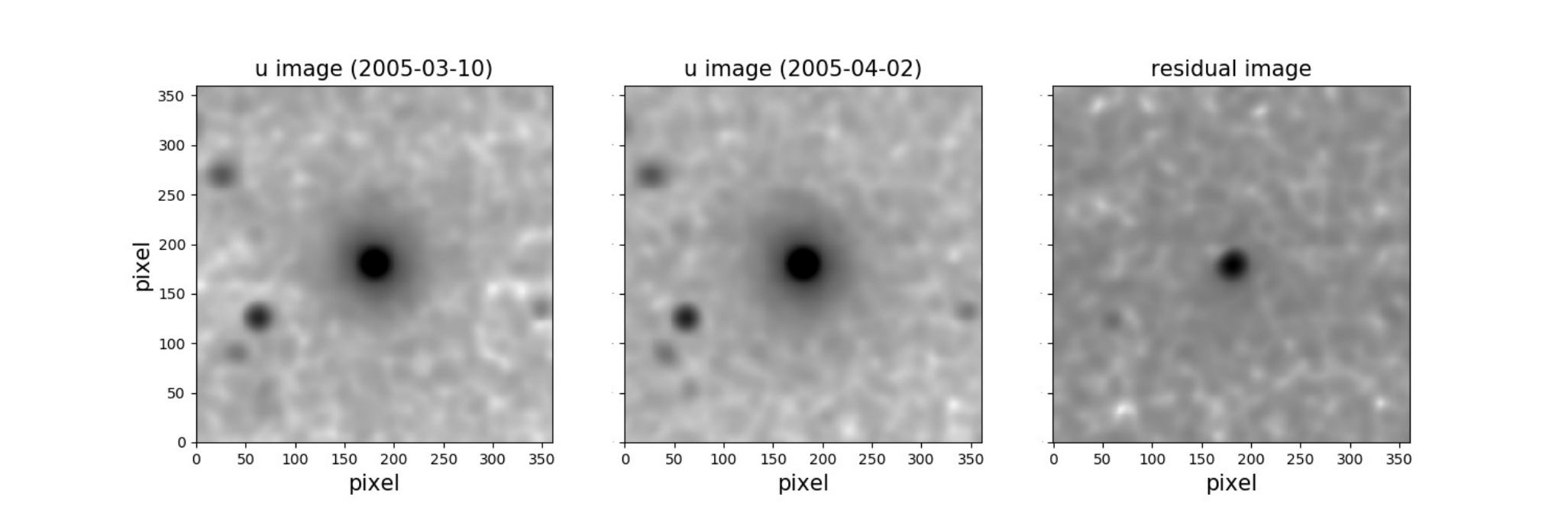}
\caption{From left to right, cutouts of SDSS $u$-band images in 2005-03 and 2005-04 observations 
and the residual image with a size of 350 $\times$ 350 pixels centered on NGC 5092. 
The two data images are convolved with a Gaussian profile with FWHM=4.45\arcsec and 4.43\arcsec, respectively, so
that the smoothed images have matching PSFs in FWHM. The residual image is obtained by subtracting the image on 2005 March 10
from that on 2005 April 2.}
\label{fig:sdss_diff}
\end{figure*}

\begin{figure*}
\vspace{-1cm}
\plotone{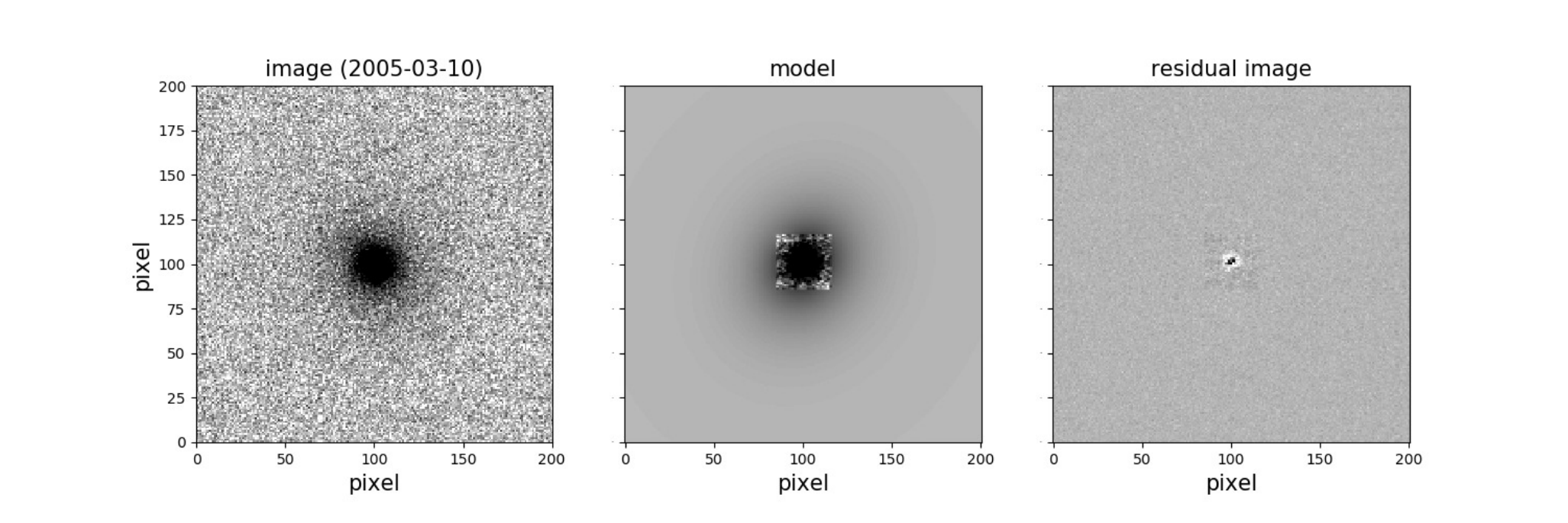}
\plotone{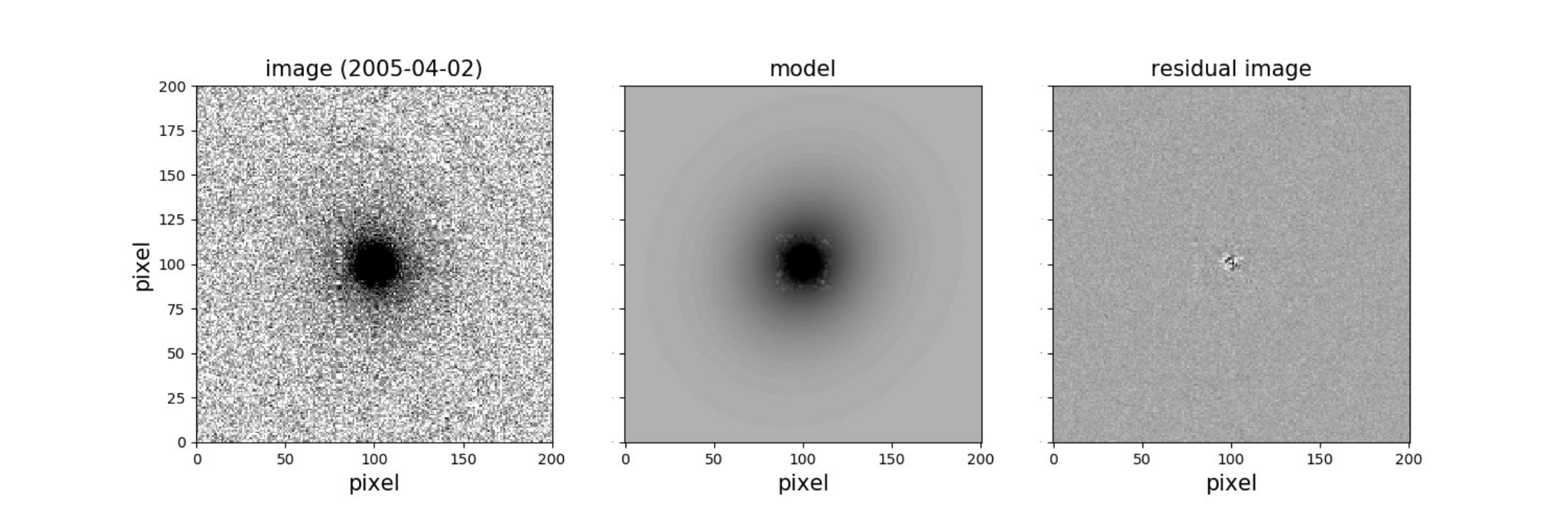}
\includegraphics[width=0.6\textwidth]{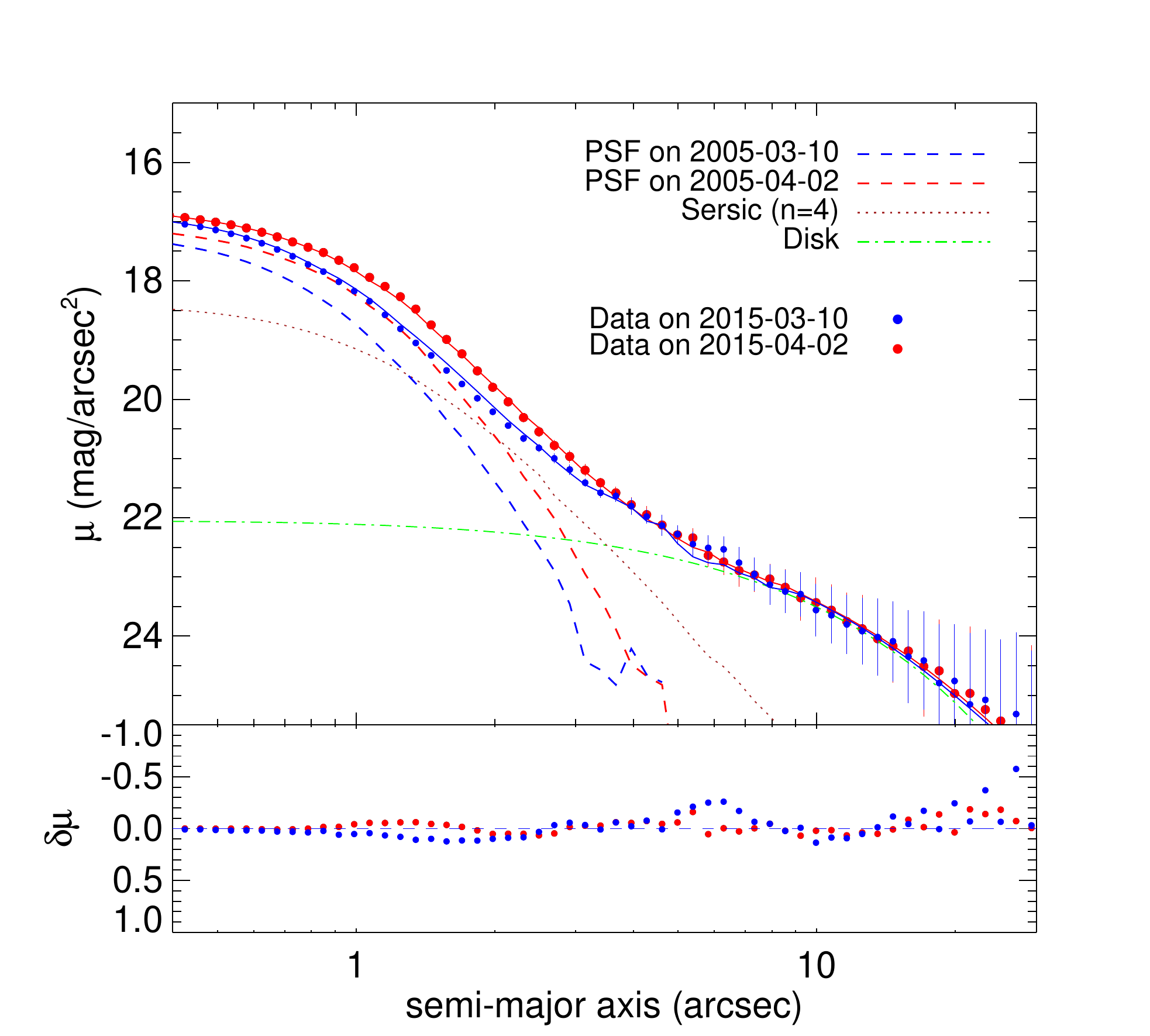}
\caption{ 
First and second row: 2D image decomposition in $u$ band of the galaxy NGC~5092 in SDSS 2005-03 and 2005-04 observations, respectively.
The left column shows the data images. The middle column shows the best-fit model: exponential disk+S$\acute{\mathrm{e}}$rsic+PSF+Sky.
The right column shows the residual images derived by subtracting the model from the original data.
Last row: sky-background-subtracted radial profiles of the data images in $u$ band on 2005 March 10 (blue circles) and 2005 April 2 (red circles). 
The individual components are PSF on 2005 March 10 (blue dashed), PSF on 2005 Apri 2 (red dashed), 
S$\acute{\mathrm{e}}$rsic (brown dotted) and disk (green dashed-dotted). The integrated models are shown with 
blue and red solid lines for the 2005-03 and 2005-04 observations, respectively. The bottom panel shows the residual distribution.}
\label{fig:sdss images}
\end{figure*}

NGC~5092 was observed for broadband photometric imaging in the $u,g,r,i,z$ bands in the SDSS
on 2005 March 10 (run 5183, camcol 6, field 441)
and 2005 April 2 (run 5224, camcol 1, field 299).
The detailed data analysis can be found in Appendix \ref{sec:app.sdss} and the obtained magnitudes of the galaxy are 
listed in Table \ref{tab:photometry}. The two SDSS imaging observations, performed near the inferred epoch of the tidal disruption, 
can also be used to examine any possible optical flare using the image subtraction method. Compared to the 2005 March 10 images, 
there is a clear excess in the core of the galaxy in the 2005 April 2 images, indicating a brightening, 
and thus a non-stellar process in the nuclear region (see Figure \ref{fig:sdss_diff}). To quantify this possible flare, 
we decompose the nuclear emission from the host galaxy for the drC images using the 2D imaging analysis tool \textsc{galfit} (\citealp{2002AJ....124..266P}). 
Figure \ref{fig:sdss images} shows, for the $u$ band as demonstration, the resultant model and residual images from the image decomposition, 
as well as the radial profiles of the surface brightness and their model components of the galaxy. 
The fitted parameters of the nuclear emission (point-spread function [PSF]) and the galactic components are summarized in Table \ref{tab:galfit summary}.
The fitted positional offset of the flare and the nucleus is only subpixel (0.396\arcsec per pixel), or less than 150 pc.
We thus consider the optical flareto have most likely originated from the central SMBH. Please refer to Appendix \ref{sec:app.sdss} for the details of the 
image subtraction and decomposition.

\begin{deluxetable*}{cccccccc}
 \tablecolumns{8} 
\tablecaption{Obtained parameters of 2D image decomposition with \textsc{galfit}}
\tablehead{& &\colhead{PSF\_mag}& \colhead{S$\acute{\mathrm{e}}$rsic\_mag}&\colhead{S$\acute{\mathrm{e}}$rsic\_r$_e$}&\colhead{Exp't\_mag}& \colhead{Exp't\_r$_s$ } \\
Obs&Filter&(mag)&(mag)&(arcsec)&(mag)&(arcsec)&$\chi^2_\nu$\\
(1)&(2)&(3)&(4)&(5)&(6)&(7)&(8)}
\startdata
\hline
5224&$u$&15.93&16.34&0.91&15.90&6.70&1.02\\
5183&$u$&16.26&...&...&...&...&1.03 \\
5224&$g$&15.97&13.98&7.58&14.82&9.12&0.45\\
5183&$g$&16.12&...&...&...&...&0.34\\
5224&$r$&16.29&12.97&9.76&14.33&9.95&0.26\\
5183&$r$&16.30&...&...&...&...&0.27 \\
5224&$i$&16.72&12.70&8.18&13.62&10.32&0.26\\
5183&$i$&16.77&...&...&...&...&0.27 \\
5224&$z$&16.88&12.42&8.34&13.48&9.27&0.65\\
5183&$z$&17.21&...&...&...&...&0.66\\
\enddata
\tablecomments {Column(1): SDSS observation ID, 5224 for observation on 2005 April 2, 5183 for observation on 2005 March 10. 
Column (2): SDSS filter. Column (3): integrated magnitude of the PSF component. Column (4): integrated magnitude of the
S$\acute{\mathrm{e}}$rsic component. Column (5): effective radius of the S$\acute{\mathrm{e}}$rsic component. Column (6): integrated magnitude of the exponential disk
component. Column (7): scale length of the exponential disk component. Column (8): best
$\chi^2_\nu$ for fitting. The magnitudes are not corrected for Galactic extinction.}
\label{tab:galfit summary}
\end{deluxetable*}


\subsubsection{XMM OM and {\it Swift} UVOT data}
During the {\it XMM-Newton} observation in 2013, the source was also observed 
with the Optical/UV Monitor (OM; \citealp{2001A&A...365L..36M}) in 
\texttt{image} mode using the $U$ filter, which is centered at 3440 {\AA}.
The observation was split into five parts with exposures of 1120s in length for each.
We present the detailed analysis in Appendix \ref{sec:app.xmm} and summarize the photometric results in Table \ref{tab:OM photometry} for each exposure.
We then take the mean 
and standard deviation of the five exposures as the aperture AB magnitude measurement of the galaxy, which is $15.67 \pm 0.01$.

\begin{deluxetable*}{ccccccc}
\tablecolumns{7} 
\tablecaption{OM aperture photometry}
\tablehead{\colhead{ExpID}&\colhead{Duration (s)}&\colhead{Source Counts}&\colhead{Source\_corr\_factor}&\colhead{Bkg Counts}&\colhead{Bkg\_corr\_factor}&\colhead{Mag}\\
(1)&(2)&(3)&(4)&(5)&(6)&(7)}
\startdata
\hline
S006& 1120 & 46094 $\pm$ 215 & 1.1 & 20437 $\pm$ 143 & 1.1 & 15.66 $\pm$ 0.01\\
S401& 1120 & 45958 $\pm$ 214 & 1.1 & 20487 $\pm$ 143 & 1.1 & 15.68 $\pm$ 0.01\\
S402& 1120 & 46262 $\pm$ 215 & 1.1 & 20959 $\pm$ 145 & 1.1 & 15.67 $\pm$ 0.01\\
S403& 1120 & 45603 $\pm$ 214 & 1.2 & 20602 $\pm$ 144 & 1.1 & 15.67 $\pm$ 0.01\\
S404& 1120 & 46049 $\pm$ 215 & 1.1 & 20562 $\pm$ 143 & 1.1 & 15.66 $\pm$ 0.01\\
\enddata
\tablecomments{ Column (1): identification of each exposure. Column (2): duration of each exposure, in unit of second. Column (3): source counts in a
circular with radius of 20\arcsec, measured by DS9 tool. Column (4): source correction factor for coincidence loss taken from the relevant exposure SWSRLI files. 
Column (5): background counts in a circular region with radius of 20\arcsec. Column (6): background correction factor for coincidence loss taken from the relevant exposure SWSRLI files. 
Column (7): magnitude calculated with the corrected source and background count rates in each exposure.}
\label{tab:OM photometry}
\end{deluxetable*}


The Ultraviolet/Optical Telescope (UVOT; \citealp{2005SSRv..120...95R}) observations were obtained in $U$ (3465 \AA) and 
UVW1 (2600 \AA) (\citealp{2008MNRAS.383..627P}) bands in \texttt{image} mode during the {\it Swift} observation in 2018. 
We measure the aperture magnitude using the UVOT software task \texttt{uvotsource} and present the results in Table \ref{tab:photometry}. 
Compared to the SDSS observation in 2005, the source was about 1 magn fainter in the $U$ band.

\subsection{{\it GALEX} UV data}

\begin {figure*}
 \centering
 \includegraphics[width=0.6\textwidth]{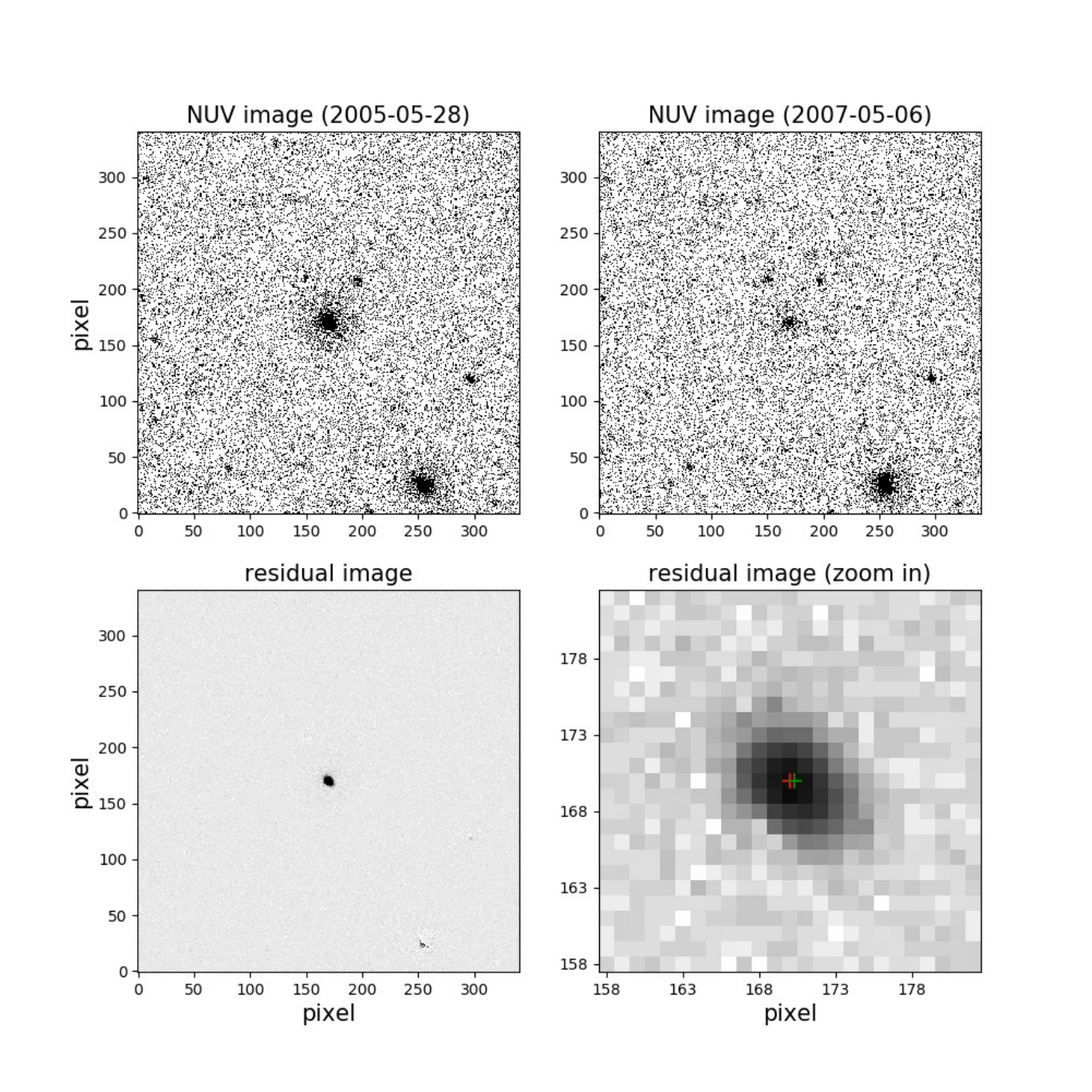}
 \caption{Top: {\it GALEX} NUV images of NGC 5092 on 2005 May 28 (left) and 2007 May 06 (right)
                     with size of 341 $\times$ 341 pixels centered on NGC 5092.
          Bottom left: residual image derived by subtracting the NUV image on 2007 May 06 from that on 2005 May 28. 
Bottom right: zoom-in version of the residual image with a size of 25 $\times$ 25 pixels. The red and green plus 
signs on the zoom-in residual image show the center of the residual image and the optical center of the galaxy, respectively.}
 \label{fig:diff_nuv}
\end{figure*}

The galaxy NGC~5092 was also imaged and detected by {\it GALEX} three times in near-UV (NUV, 1771--2831\,\AA) on 2005 May 28, 
2005 June 11, and 2007 May 6 and once in FUV (1344--1786\,\AA) on 2007 May 6.
The NUV imaging observations, performed from just a few months to two years after the tidal disruption,
provide valuable data to examine any UV radiation of the nucleus possibly associated with the X-ray flare.
We perform image subtraction on the calibrated NUV images between the 2005 May and 2007 observations and 
find the nuclear emission to be much brighter in 2005 than in 2007 (see Figure \ref{fig:diff_nuv}), 
indicating an NUV flare originating from the activity of the SMBH, likely associated with the 
X-ray flare. To quantify the amplitude of the variability, the fluxes of the entire galaxy are measured from 
the reduced images by using gPhoton (\citealp {2016ApJ...833..292M}) tool gAperture, with a photometry aperture of 20\arcsec. 
The aperture magnitudes for all the three NUV observations and the one FUV observations are presented in Table \ref{tab:photometry}. 
Compared to the 2007 observation, the source was brighter by $\sim$\,3 mag in the NUV band in 2005.
Details of the aperture photometry and image subtraction are described in Appendix \ref{sec:app.galex}.
\subsection{UV-Optical flare}
\label{sec:optical/uv flare}
\begin{figure*}
 \centering
 \includegraphics[width=0.7\textwidth]{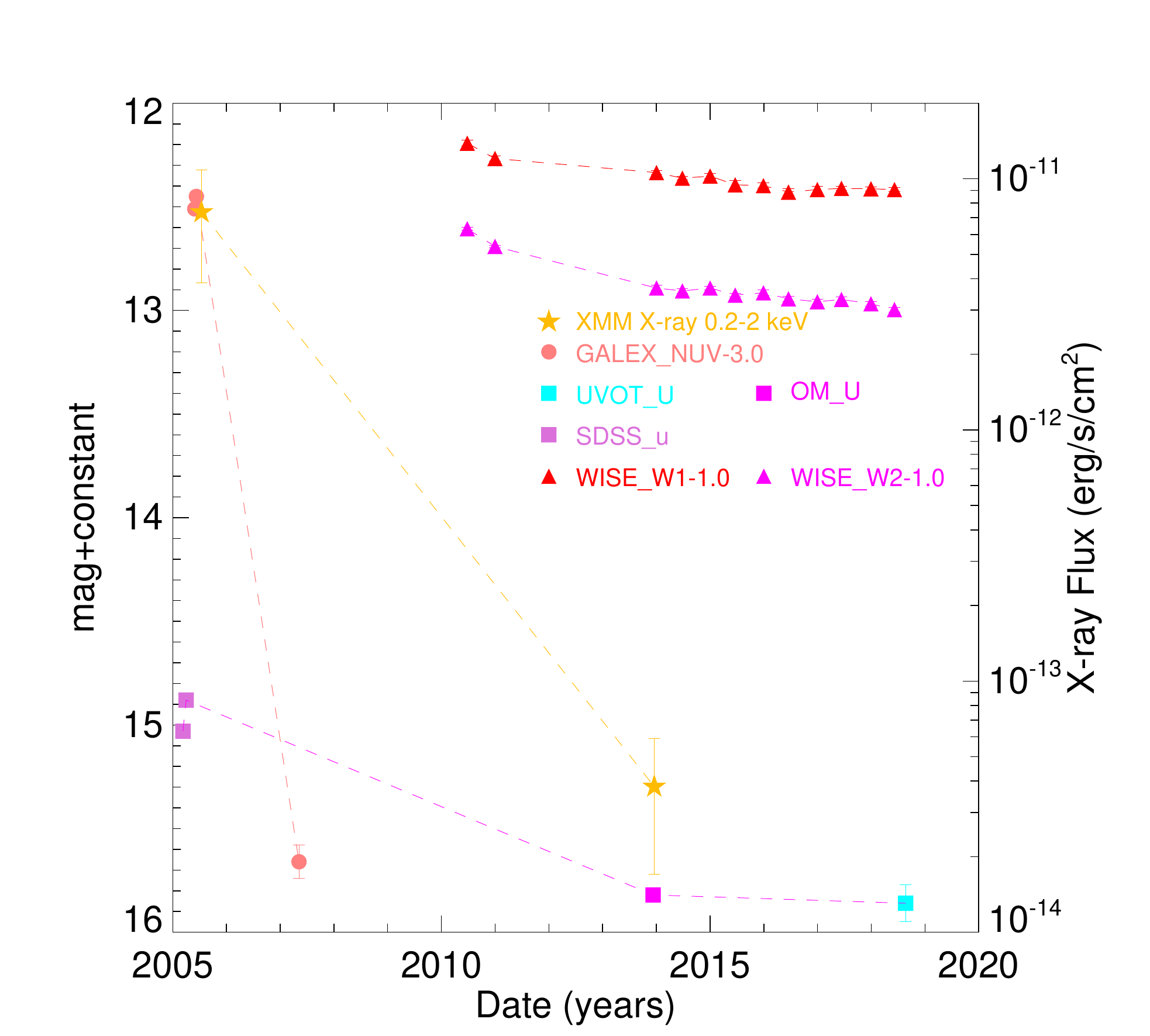}
 \caption{Multiwavelength photometric results of the entire galaxy NGC 5092 from IR to X-ray. All IR to UV magnitudes 
are in units of the AB magnitude system (left scale), and X-ray 0.2--2\,keV fluxes are shown in erg\,s$^{-1}$\,cm$^{-2}$. Note that the data are not corrected 
for Galactic extinction and error bars are in some cases smaller than 
the points. The dashed lines are used to connect the data points in the same filter and do not trace the decline law.}
 \label{fig:multi lc}
\end{figure*}

\begin{figure*}
 \centering
 \includegraphics[width=0.6\textwidth]{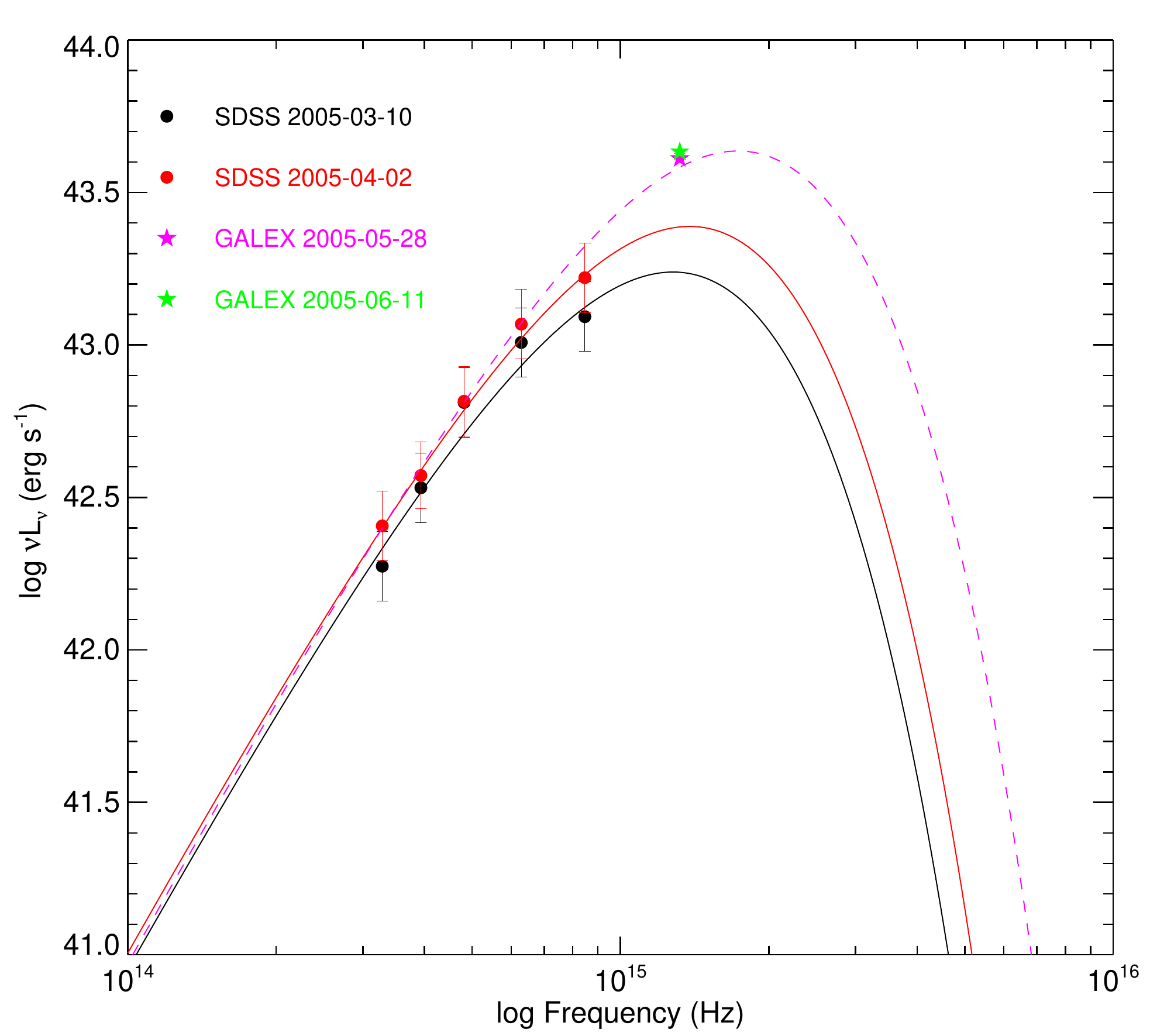}
 \caption{Galactic-extinction-corrected optical--UV spectra of XMMSL1 J1319+2259. The black
and red circles represent the PSF fluxes derived from 2D image decomposition of SDSS 2005 March 10 and 2005 April 02 images with \textsc{galfit}, respectively. 
The black and red solid lines are the best-fit blackbody curves for these two observations, respectively.
The purple and green stars are {\it GALEX} NUV data on 2005 May 28 and 2005 Jun 11, respectively, after subtracting the host galaxy contribution.
The purple dashed line shows the best-fit blackbody using the SDSS data on 2005 April 02 and {\it GALEX} NUV data on 2005 May 28.}
 \label{fig:broad band sed}
\end{figure*}

Figure \ref{fig:multi lc} shows a summary of the photometric results of the galaxy NGC~5092 obtained above taken in 
various optical and UV bands (without Galactic extinction correction).
Clearly, the {\it GALEX} NUV and optical $u$($U$) fluxes dropped significantly by about 3 and 1 mag, 
respectively,\footnote{Although the bandpasses of three filters SDSS-$u$, OM-U and UVOT-U are not the same, 
we find that differences among their measured magnitudes are less than 0.2 by comparing stars in the fields 
of view of the observations.} on timescales of a few years or longer.
Now we study the quasi-simultaneous optical-UV spectra of the flare, which were taken 1--2 months apart. 
The optical luminosities of the galactic nucleus are calculated at the central frequencies of the five SDSS filters.
To do this, the PSF fluxes derived from the 2D image decomposition in Section \ref{sdss image} are used for the 2005-03 
and 2005-04 observations, which are assumed to be dominated by the flare emission. The resulting optical spectra of the flare at the two epochs are
shown in Figure \ref{fig:broad band sed}. The two spectra can be well fitted with a blackbody model with 
temperatures of 1.6$^{+1.3}_{-0.5} \times 10^4$\,K (2005-03) and 1.7$^{+1.9}_{-0.6} \times 10^4$\,K (2005-04) (90\% uncertainties), 
respectively. The relatively large uncertainties are due to the fact that the peak or turnover of the spectrum is not sampled by observations.
We also calculate the ultraviolet luminosity of the flare from the {\it GALEX} NUV photometry by subtracting the galaxy contribution, 
which is estimated from the SED fitting of the galaxy starlight (see Section \ref{diss:host galaxy}). 
The luminosities at 2305~\AA \ for the 2005-05 and 2005-06 observations are overplotted in Figure \ref{fig:broad band sed}. 
Assuming that the luminosities did not vary significantly between 2005-04 and 2005-05, we fit jointly 
the optical and NUV luminosities with a single blackbody, resulting in a 
temperature of 2.4$^{+0.6}_{-0.3}$\,$\times\,10^4$\,K. A powerlaw can also give a good fit to the data, 
with $\Gamma$\,$\sim$\,0.2. Thus, the observed optical--UV flare emission around March--April, 
if assumed to be a blackbody, had a temperature of $\sim$\,(1--2)\,$\times 10^4$\,K, consistent with those in previously 
reported optical TDEs (\citealp{2018ApJ...852...72V}). The integrated optical--UV luminosities, from 100 to 10000~{\angstrom}, 
of the flare are estimated to be around $\sim$\,(2--6)\,$\times 10^{43}$\,erg\,s$^{-1} $, corresponding to an Eddington ratio of $\sim$\,(3--9)\,$\times 10^{-3}$.

We note that the possible existence of a weak, AGN-like activity in the galaxy cannot be ruled out (see Section \ref{sec:optical spectra}), 
which might contribute to the PSF fluxes. While in $u$ band, the whole galaxy in 2005 April was about 1 mag brighter than that 
in the UVOT observation in 2018, suggesting that the PSF flux in at least the $u$-band must be dominated by the flare. 
The blackbody temperature calculated by using only the $u$ band PSF flux in 2005 April and
the NUV in 2005 May is $\sim$\,35,000\,K, also in the range of those previously reported TDEs.

\subsection{WISE and 2MASS}
\label{sec:IR}

In the MIR band, NGC~5092 was bright enough to be detected in a single exposure with {\it WISE} in the $W1$ (3.4~$\mu$m) and $W2$ (4.6~$\mu$m) bands. 
We analyzed the {\it WISE} observational data for NGC~5092, which were taken from 2010 to 2018, and extracted light curves with 12 time bins in the two wavebands. 
The details of the data analysis are described in Appendix \ref{sec:app.wise}.
The MIR light curves are overplotted in Figure \ref {fig:multi lc}, which show a trend of fading over a time span of about 6 yr,
and appear to remain mostly constant in the observations of the last several epochs. To estimate the contribution from the host galaxy, 
we consider the last five epochs for $W1$ and the last one epoch for $W2$ as the quiescent state, 
which have averaged magnitudes of $\rm W_1 = 13.73 \pm 0.01$ and $\rm W_2 = 13.68 \pm 0.01$. 
Assuming that the emission in the $W1$ and $W2$ bands originates from a blackbody, we calculate the temperatures in each epoch using
the host-galaxy-substracted fluxes. We find that the temperatures spread in the range of $\sim$\,600--1500\,K,
consistent with those of the previously reported TDE IR echoes (\citealp{2009ApJ...701..105K}; \citealp{2016ApJ...832..188D, 2017ApJ...841L...8D}; 
\citealp{2016ApJ...828L..14J,2017ApJ...850...63J}; \citealp{2016ApJ...829...19V, 2018MNRAS.477.2943W}).
We also obtained the NIR $JHK_{\rm S}$ images from 2MASS and measured the aperture magnitudes as shown in Table \ref{tab:photometry}.

\section{Optical spectral analysis}
\label{sec:optical spectra} 
\begin{figure*}
 \centering
 \includegraphics[width=0.6\textwidth]{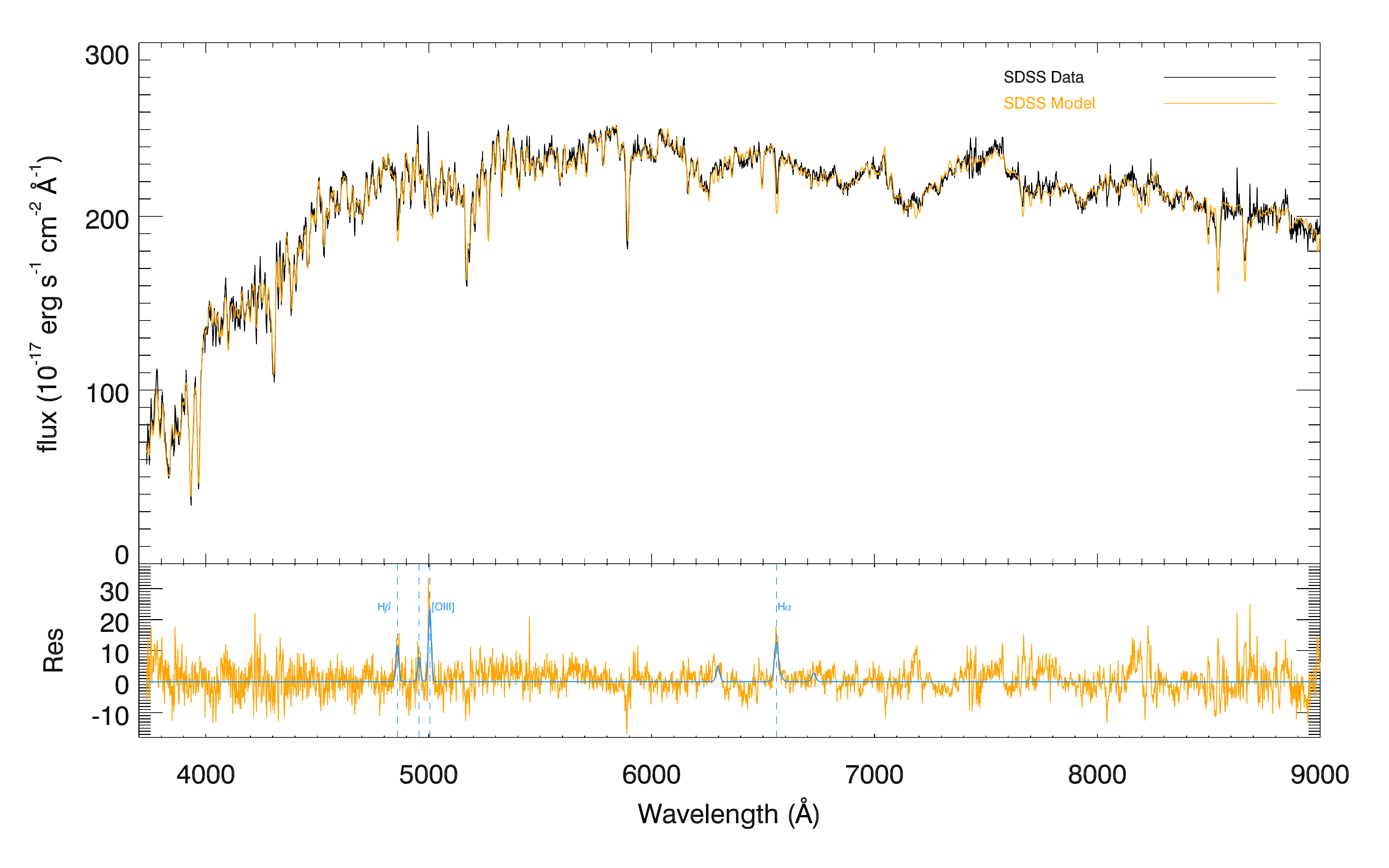}
 \caption{Optical spectrum of NGC 5092 obtained in SDSS on 2008 Feb 11. Flux calibrated data are shown in black, 
and the stellar population model is shown in orange. The starlight-model-subtracted spectrum is shown in orange in the bottom panel.}
 \label{fig:host optical spectrum}
\end{figure*}

An optical spectrum of NGC~5092 was obtained by the SDSS on 2008 February 11, which is about 3 yr after the optical/UV flare. 
The fiber was centered at the nucleus of the galaxy with a radius of 3\arcsec.  Figure \ref{fig:host optical spectrum} shows the spectrum of NGC~5092, 
which is dominated by host starlight. There seem to be no apparent emission lines in this spectrum by visual check. Hence, 
it necessitates proper model and subtraction of the stellar continuum to examine possible weak emission-line features. 
Generally, we conducted the continuum fitting following the procedures described in \cite{2006ApJS..166..128Z}; 
for details see Appendix \ref{sec:app.starlight}. As shown in figure \ref{fig:host optical spectrum}, 
the galaxy starlight model (orange) gives a reasonable fit to the optical spectrum (black), with a dominant 
middle-aged stellar population of 2.5\,Gyr. 
The best-fit stellar velocity dispersion is $\sigma^*=156\pm 26$\,km \,s$^{-1}$
(corrected for instrumental broadening of 56 km s$^{-1}$ as measured from arc spectra and tabulated by the SDSS). 
We can thus estimate the mass of the central BH of the galaxy from $\sigma^*$ using the formalism of \citealp{2005SSRv..116..523F}:
\begin{equation}
  M_{BH}=1.66(\pm 0.24) \times 10^{8} {\rm M}_{\odot} \left (\frac{\sigma^*}{200 \rm km \rm s^{-1}} \right )^{4.86 (\pm 0.43)},
\end{equation}
yielding  $M_{\rm BH}=5.0^{+5.6}_{-2.9} \times 10^{7} {\rm M}_{\odot}$.

After removing the continuum, several narrow emission lines, including {\oiii} and Balmer lines, are found in the residual spectrum, 
and no broad-line features are detected. The fitting methods are detailed in the Appendix \ref{sec:app.emission}. 
Among these emission lines, {\oiii}$\lambda 5007$ and {\ha} are detected with a signal-to-noise ratio (S/N) of \,>\,3, 
and {\hb} is marginally detected with S/N\,$=$\,2.4. Note that, apart from the statistical uncertainty in the line fitting, 
the uncertainty introduced by continuum subtraction has also been considered in the estimation of S/N; 
for details see Appendix \ref{sec:app.emission}. The line measurements, including the fluxes and equivalent widths (EWs), 
are summarized in Table \ref{tab:emission line}, and the EWs are consistent with those given by MPA-JHU 
catalogs\footnote {\url{https://www.sdss.org/dr14/spectro/galaxy_mpajhu/}} based on the methods of \citet{2004MNRAS.351.1151B}, 
\citet{2003MNRAS.346.1055K}, and \citet{2004ApJ...613..898T}, e.g., $1.8 \pm 0.4$ versus $2.0 \pm 0.1$ for EW ({\oiii}$\lambda5007$), 
$1.1 \pm 0.3$ versus $0.8 \pm 0.1$ for EW ({\ha}).

\begin{deluxetable*}{ccccccc}
\tablecolumns{7} 
\tablecaption{Optical Emission-line measurements}
\tablehead{ & \colhead{{\oiii}$\lambda5007$}&\colhead{{\ha}}&\colhead{{\hb}}&\colhead{{\nii}$\lambda6583$}&\colhead{{\sii}$\lambda6717$}&\colhead{{\sii}$\lambda6731$}}
\startdata
\hline
Flux (10$^{-17}$ erg s$^{-1}$ cm$^{-2}$)$^{\rm a}$&382.4 $\pm$ 78.8&268.5 $\pm$ 78.8&186.2 $\pm$ 78.8& $<$ 94.2& $<$ 54.2& $<$ 126.0\\
\hline
Equivalent width (\AA)$^{\rm b}$&1.8 $\pm$ 0.4&1.1 $\pm$ 0.3&0.9 $\pm$ 0.4& $>$ 0.4& $>$ 0.2& $>$ 0.6\\
\enddata
\tablecomments{ $^{\rm a}$ Fluxes or 90\% upper limits of the emission lines after subtracting the 
starlight (please refer to Appendix \ref{sec:app.emission} for the details of the measurements). 
$^{\rm b}$ Equivalent widths (EWs) of the emission lines after correcting stellar absorption.}
\label{tab:emission line}
\end{deluxetable*}

\begin{figure*}
 \centering
 \includegraphics[width=0.6\textwidth]{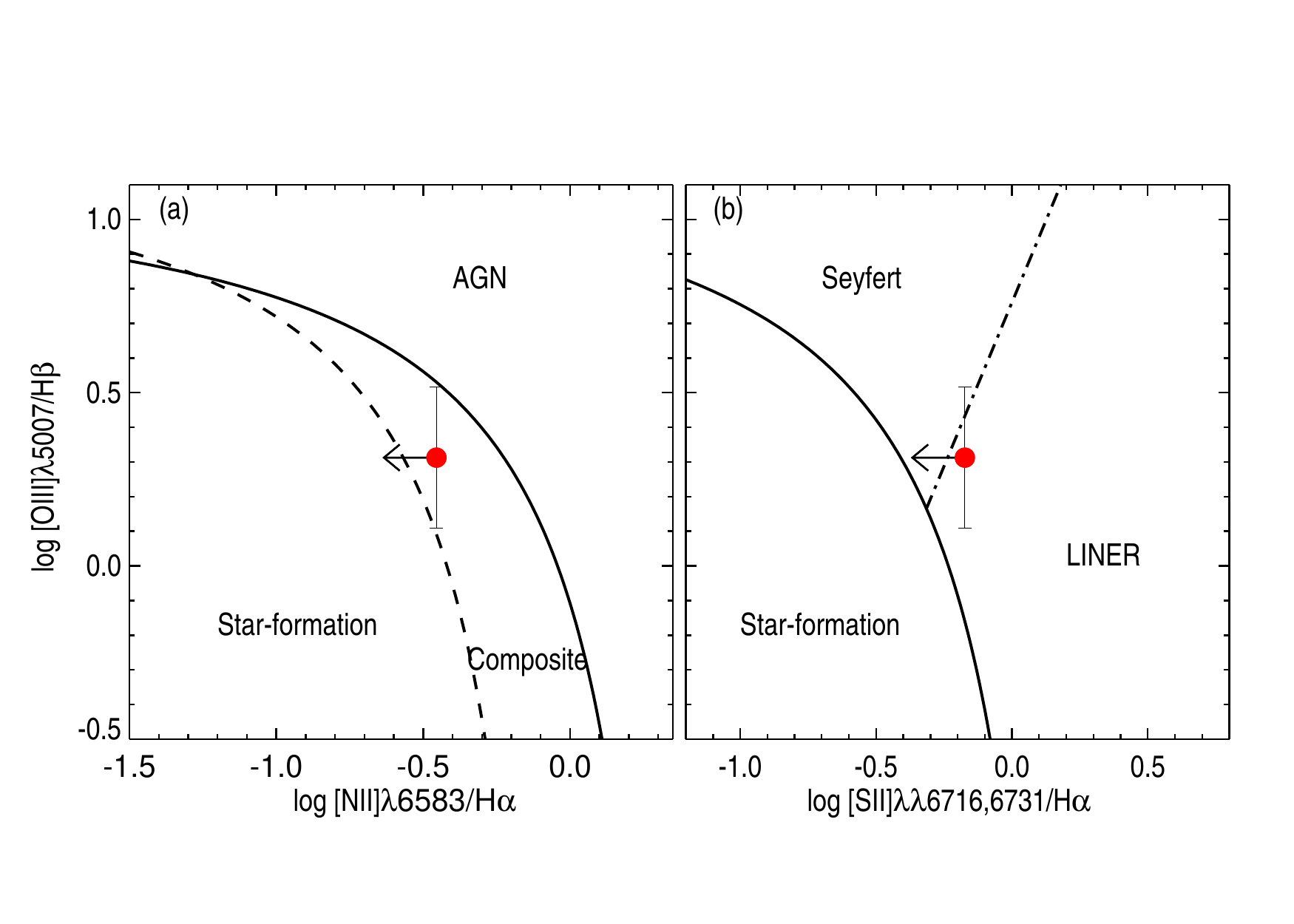}
 \caption{Location of NGC 5092 on the BPT diagram of {\oiii}$\lambda5007$/ {\hb} $vs.$ 
{\nii}$\lambda6583$/{\ha} and {\oiii}$\lambda5007$/ {\ha} $vs.$ {\sii}/{\hb}.
In panel (a),  
the solid and dashed curves are respectively the theoretical line for extreme starburst (\citealp{2001ApJ...556..121K}) 
and the empirical dividing line between starformation and Seyferts or LINER, respectively (\citealp{2003MNRAS.341...33K}). 
Sources that lie between the two lines are classified as composite starburst/AGN objects. In panel (b), 
the solid curve has the same meaning as in panel (a), and the dashed-dotted curve is the Seyfert-LINER line taken from \citet{2006MNRAS.372..961K}.}
 \label{fig:bpt 5092}
\end{figure*}

The narrow-line diagnostic diagram, the so-called BPT diagram (\citealp{1981PASP...93....5B}), 
is used to investigate the mechanism of the line emission. As {\nii}$\lambda\lambda 6548,6583$ and {\sii}$\lambda\lambda 6716, 6731$
are not robustly detected, their flux upper limits (see Appendix \ref{sec:app.upperlimit} for details) are adopted to calculate the line ratios.
The locations of NGC~5092 on the diagram of {\oiii}$\lambda5007$/ {\hb} $versus$ 
{\nii}$\lambda6583$/H$\alpha$ and {\oiii}$\lambda5007$/ {\hb} $versus$ 
{\nii}/{\ha} are shown in Figure~\ref{fig:bpt 5092}.
As can be seen, the location of the line ratios of NGC~5092 is close to the boundaries among the AGN, 
low-ionization nuclear emission-line region (LINER), and star-formation regions. Given the relatively large 
uncertainties of the line ratios due to the weakness of the lines, a reliable classification of emission line mechanism is hard to achieve, 
though the {\oiii}/{\hb}-{\nii}/{\ha} diagram may hint at LINER.

\section{Discussion}
\label{discussion}
\subsection {XMMSL1 J1319+2259: A TDE in NGC~5092}

The X-ray source XMMSL1~J1319+2259 with a 0.2--2\,keV luminosity of $\sim$\,$9 \times 10^{42}$\,erg\,s$^{-1}$ as detected in the XMMSL in 2005 July, 
brightened by a factor of $> 20$ as compared with the previous RASS upper limit, and faded to a flux $\sim$\,240 times fainter when it was reobserved 8\,yr later. 
The X-ray spectrum of the flare is very soft and can be described with a blackbody at $kT\sim$\,60\,eV, 
with an overall luminosity of $\sim$\,$1.5 \times 10^{43}$\,erg\,s$^{-1}$. The features are hardly explained by a supernova but 
resemble those of X-ray TDEs reported previously. Assuming a t$^{-5/3}$ law of flux decline characteristic of TDEs, 
the disruption time is derived to be around 2005 February to April, and the 0.2--2\,keV peak luminosity around (1--7)\,$\times 10^{43}$\,erg\,s$^{-1}$, 
assuming that the flux began to decrease from 1 to 3 months after the disruption. Interestingly enough, an optical--UV flare is also found 
to occur at the galactic nucleus around the epoch of the X-ray flux, suggesting their physical association.
The optical--UV flare can be fitted with blackbody radiation at a temperature of $\sim$\,(1--2)$\times 10^4$\,K. 
Assuming a spherical geometry, following the Stefan--Boltzmann law, the sizes of the optical/UV and X-ray emission regions 
are around 100 $R_{\rm g}$ and several $R_{\rm g}$, respectively, where $R_{\rm g}$ is the gravitational radius ($R_{\rm g}=\frac{GM_{\rm BH}}{c^2}$),
for the derived BH mass of $\sim$\,$5 \times 10^{7} \rm M_{\odot}$ from the $M_{\rm BH}$--$\sigma^*$ relationship. 
Besides, the continuous decline in the MIR fluxes of the galaxy in the {\it WISE} $W1$ and $W2$ bands from 2010 to 2018 
can be naturally attributed to the dust-reprocessed light of the flare at higher energy bands 
(\citealp{2016ApJ...828L..14J,2016MNRAS.458..575L, 2016ApJ...829...19V}).
All these features are characteristic of TDEs observed previously, which we consider to be the most likely interpretation for this object, 
albeit the sparse data sampling.

The X-ray emission as observed with {\it XMM-Newton} in 2013 had a much lower luminosity of $\sim$\,$3 \times 10^{40}$\,erg\,s$^{-1}$ 
and the spectrum became much harder with $\Gamma=2.1$. At such a low luminosity level, the contribution from the host galaxy may not be negligible. 
Assuming that only 10\% of the observed flux comes from the TDE process, the derived tidal disruption time would be around 2006-06 and the 
derived peak luminosity in 0.2--2\,keV is $\sim$\,$10^{43}$\,erg\,s$^{-1}$.
Although the derived disruption time might be later than the optical/UV brightening, we note that it is subject to large uncertainty, 
as it depends on the estimation of the flux and the assumed decline law.
XMMSL1 J1319+2259 is a rare example of a TDE candidate that shows contemporaneous X-ray and optical
flares, while the majority of observed TDEs were detected either in optical/UV or in X-ray. According to \citet{2018ApJ...859L..20D}, the diversity could be 
interpreted by a unified model where the spectral properties of the TDE depend mainly on the viewing angle with respect to the orientation of the disk. 

\subsection{AGN variability?}

Although all the observed properties of XMMSL1~J1319+2259 and its associated multiwavelength flares are typical of the previously reported TDEs, 
the possibility of AGN variability may not be ruled out completely, due to the possible 
existence of weak nuclear activity in the host galaxy, as revealed from the ratios of the weak 
emission lines shown in the optical spectrum of the galaxy. Radio emission, which is commonly used to be an important diagnostic of AGN activity, 
is not detected in NGC~5092 either in the Faint Images of the Radio Sky at Twenty cm (FIRST; \citealp{1997ApJ...475..479W}) or in the 
NRAO VLA Sky Survey (\citealp{1998AJ....115.1693C}) database, with the upper limit given by FIRST to be 0.2\,mJy. 
We note that the X-ray emission of NGC~5092 at the nonflaring states is rather weak, e.g. $L_{0.2-2~\rm keV} \sim 3 \times 10^{40}$\,erg\,s$^{-1}$ 
from the {\it XMM-Newton} pointed observation in 2013. Such a luminosity is comparable to typical X-ray emission from inactive galaxies, 
which can be estimated from the  NIR fluxes (2MASS $H$ bands) to be $L_{X} \sim 6 \times 10^{39}$\,erg\,s$^{-1}$, with a standard deviation of 0.5 dex, 
using the X-ray--NIR luminosity relation for early-type galaxies (\citealp{2006MNRAS.367..627E}).
This sets a strong constraint that even if there exists a nuclear activity in NGC~5092, it has to be rather weak. 
This is also supported by the $W1-W2$ color from the {\it WISE} measurements, which indicates the lack of a strong AGN  in NGC~5092.

However, analysis of the optical spectrum taken in 2008, about 3 yr after the disruption event, shows that NGC~5092 may be 
located in the Seyfert/LINER regime in the BPT diagram, although the classification is rather uncertain owing to the large 
uncertainties in the measured line ratios. Assuming that the narrow emission lines, {\oiii} and {\ha}, are from the persistent weak AGN radiation, 
the expected AGN luminosity is $L_{0.2-2~\rm keV} \sim 6 \times 10^{40}$\,erg\,s$^{-1}$, comparable to that measured with the {\it XMM-Newton} observation in 2013, 
based on the relation between the X-ray luminosities and optical emission lines for local AGNs\footnote{Please note that although the correlations 
between X-ray and optical emission lines are of relatively large scatter, the luminosity in the {\it XMM-Newton} pointed observation in 2013 is 
comparable to that expected from the best-fit correlations.} (\citealp{2006A&A...455..173P}). Thus, the possibility cannot be ruled out that this galaxy hosts a weak AGN.
However, the very soft X-ray spectrum at the flaring state ($\Gamma \sim 4$) as well as the spectral hardening with time, is typical of TDEs rather than AGNs. 
Furthermore, those weak emission lines ({\oiii} and {\ha}) are found to be relatively broader (FWHM$\sim$900 km s$^{-1}$) than those of most of AGNs (\citealp{2018ApJ...856..171Z}), 
which may be explained as originating from the TDE process. In this work we argue that the observed flare in the X-ray and multiple wavebands originated most likely from a TDE, 
although an extreme AGN variability event of a possible weakly active nucleus in NGC~5092 cannot be ruled out. 

\subsection{Infrared echo}
The {\it WISE} observations show a continuous decline in the MIR light curves as described in Section \ref{sec:IR}.
Similar flaring and fading in the MIR light curves
have also been found in several TDEs (\citealp{2016ApJ...832..188D, 2017ApJ...841L...8D}; \citealp{2016ApJ...828L..14J, 2017ApJ...850...63J}; 
\citealp{2016ApJ...829...19V, 2018MNRAS.477.2943W}), which is explained as emission from dust in the nuclear region by reprocessing the high-energy radiation of the TDE.
Such an echo radiation lags the primary radiation by a delay time of approximately $R/c$, assuming the central symmetric geometry of the irradiated dust, with a distance
$R$ to the central source and where $c$ is the speed of light. The echo luminosity depends on the bolometric luminosity of the primary radiation and the covering factor of the dust 
subtended to the central source.
\begin{figure*}
 \centering
 \includegraphics[width=0.6\textwidth]{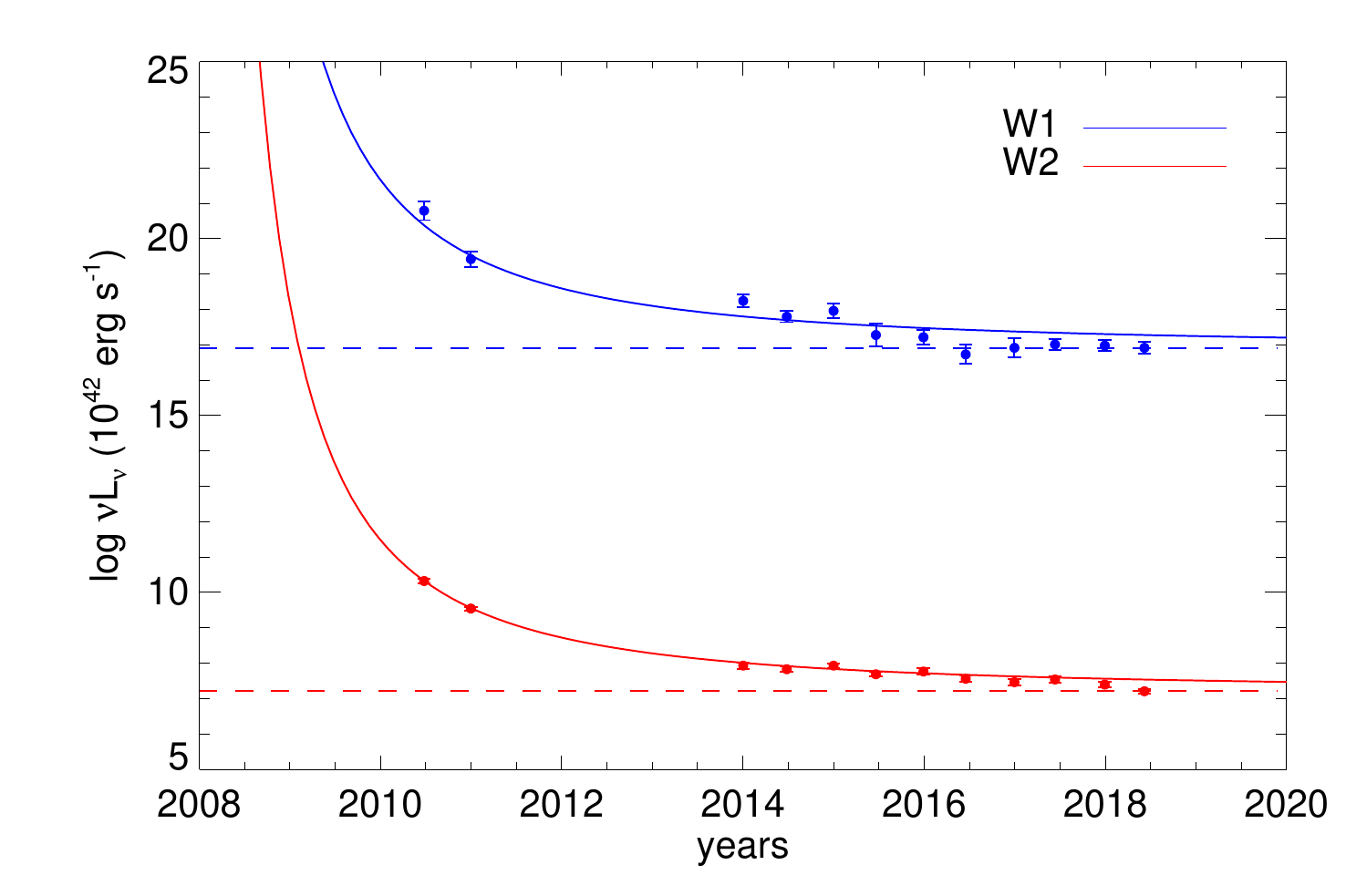}
 \caption{Smoothed $W1$ and $W2$ light curves of NGC~5092 and the dust reradiation model. The $W1$ and $W2$ data are shown in 
blue and red points with error bars, respectively. Assuming a t$^{-5/3}$ decline law and a constant dust temperature of 1000~K, the reproduced $W1$ and $W2$ light curves derived 
from the dust reradiation model are shown with blue and red curves, respectively. Also shown in dashed lines are the contributions from the host galaxy in each band.}
 \label{fig: wise_lc}
\end{figure*}

We consider the MIR light curves observed in NGC~5092 with {\it WISE} as the fading echo of dust irradiation of the TDE flaring, and we attempt to reproduce them with a simple model.
We assume that the bolometric luminosity of the TDE $L_{\rm TDE}$ decreases with time as $t^{-5/3}$ as the X-ray luminosity $L_X$ does, and the energy absorbed by the dust is
completely reradiated away.
The IR luminosity $L_{\rm IR}$ is calculated as $L_{\rm TDE}$ multiplied by the covering factor of the irradiated dust $C_f$, with a 
time lag $t_d$, 
\begin{equation}
L_{\rm IR}(t) = L_{\rm TDE}(t-t_d)*C_f = L_X(t-t_d) \times \frac{L_{\rm TDE}}{L_X} \times C_f,
\end{equation}
where $L_{\rm TDE}/L_X$ is the bolometric correction factor in the X-ray band of the TDE and is assumed to be constant over time.
The luminosities in the $W1$ and $W2$ bands can be calculated by assuming the dust emission to be blackbody with an effective temperature
$T_{\rm dust}$. As described in Section \ref{sec:IR}, the temperatures are found to lie in the range of $600-1500$\,K.
As the first-order approximation, we try to reproduce the light curves in the W1 and W2 bands by fitting them jointly with a simple model assuming a constant temperature of 1000~K. We calculate the monochromatic luminosity in the $W1$ and $W2$ bands as 
\begin{equation}
\label{eq.dust}
\nu_i L_{\nu_i}(t) = L_{\rm IR}(t) / n_i + L_{0i}, \ (i=1,2 {\rm \ for} \ W1 {\rm \ and} \ W2 {\rm \ band}) 
\end{equation}  
where $L_{0i}$ is the luminosity of the host galaxy and $L_{01} =1.6 \times 10^{43}$\,erg\,s$^{-1}$ for $W1$ and $L_{02} =7.2 \times 10^{42}$\,erg\,s$^{-1}$ 
for $W2$ (see Section \ref{sec:IR}), and $n_i$ is the correction factor at the frequency $\nu_i$ at the temperature $T_{\rm dust}$. 
For $T_{\rm dust} = 1000$\,K, the ratio between $n_1$ and $n_2$ is $n_1/n_2 = $0.9.
From Eq. \ref{eq.dust} we have
\begin{equation}
\nu_i L_{\nu_i}(t) = A_i \times L_X(t-t_d) + L_{0i}, \qquad (i=1,2) 
\end{equation} 
where the time lag $t_d$ and $A_i = \frac{L_{\rm TDE}}{L_X} \times C_f / n_i$ are both free parameters (with $A_2/A_1 = 0.9$).
We fit the model to the $W1$ and $W2$ data jointly by minimizing $\chi^2$.
The best-fit parameters are $t_d =2.5 \pm 0.1$\,yr, $A_2 = 15.1 \pm 1.1$, and the fitting result is plotted in Figure \ref {fig: wise_lc}.
Although the model is oversimplified, it can well reproduce the observed light curves of both $W1$ and $W2$, indicating that 
the observed MIR flare may be explained by the reradiation of TDE primary photons by dust. To obtain a better constraint on 
the dust distribution and richness around the BH, well-covered light curves and accurate measurement of the TDE SED are needed.

\subsection {Host galaxy}
\label{diss:host galaxy}

We perform host galaxy SED fitting with fluxes in $ugriz$ bands derived by subtracting the PSF fluxes from the aperture fluxes of the SDSS 2005-04 observation.
The fluxes are corrected for the Galactic extinction by employing the IDL code fm\_unred.pro with $E(B-V)=0.014$ (\citealp{1998ApJ...500..525S}), 
assuming the Milky Way (MW) extinction curve (\citealp{2003ApJ...594..279G}).
The code we use for galaxy SED modeling is model package Multi-wavelength Analysis of Galaxy Physical Properties (\textsc{magphys}),
a self-contained, user-friendly model package to interpret the observed SED
of galaxies in terms of galaxy-wide physical parameters pertaining the stars and the interstellar medium, 
following the approach described in \citet{2008MNRAS.388.1595D}.

\begin{figure*}
\centering
\includegraphics[width=0.6\textwidth]{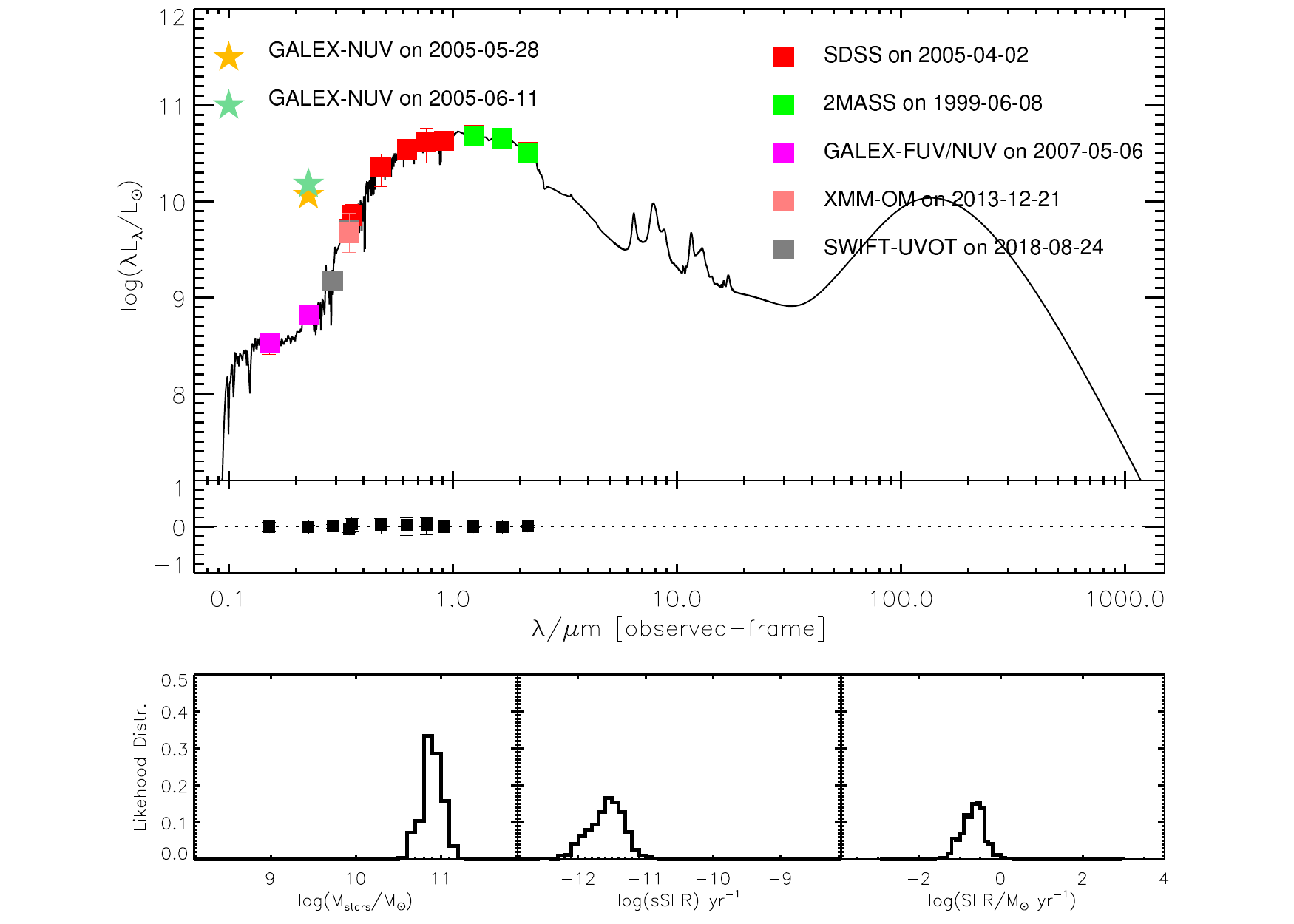}
\caption{Host galaxy SED fitting with \textsc{magphys} --- luminosity $vs.$ wavelength in logarithmic scale. 
Data points used in the fitting include the {\it GALEX} FUV and NUV in 2007, OM in 2013, UVOT in 2018, 2MASS in 1999, 
and galaxy fluxes derived from the SDSS 2005 April 02 images using \textsc{galfit}. Also shown in stars are
the Galactic-extinction-corrected {\it GALEX} NUV fluxes in 2005. The black curve represents the best fit to
the observed points.
The lower part of the figure gives the likelihood distribution of the parameters, stellar mass of the galaxy ($\rm M_{\rm stars}$),
specific star formation rate (sSFR), and star formation rate (SFR).}
\label{fig:host sed}
\end{figure*}

The galaxy SED thus obtained using the SDSS imaging photometric data only is compared with the measurements in the other bands analyzed 
in Section \ref{data reduction}. It is found that all the measurements match the extrapolation of the galaxy SED well except for the 
two {\it GALEX} NUV fluxes in 2005, which clearly show excesses as contributed by the nuclear flare. 
We thus repeat the SED fitting using all the photometric data available for the galaxy except for the {\it GALEX} NUV ones. 
This leads to a good fit, as shown in Figure \ref{fig:host sed}. The fitted stellar mass $M_{\rm stars} =7.5^{+2.3}_{-1.7} \times 10^{10}$~$\rm M_{\odot}$ (68\% conf.), 
and star formation rate $SFR=0.2 \pm 0.1 \rm M_{\odot} \rm yr^{-1}$ (68\% conf.).
The SED modeling shows that there are five random bursts experienced by the galaxy, with the last burst of 
star formation ending 1.89\,Gyr ago.
With the relationship between black hole mass and bulge $K_{\rm s}$-band luminosity (\citealp{2003ApJ...589L..21M}), we obtain
a BH mass of $1.06\times 10^{7} \rm M_{\odot}$ using the point-source 2MASS magnitude of $m_{K_{\rm s}}=11.51$, with 
a systematic error of 0.3 dex ($2\times 10^{7} \rm M_{\odot}$), which is consistent with that derived from the 
$M_{\rm BH}-\sigma^*$ relationship. The optical spectral fitting shows that the starlight of NGC~5092 is dominated by a middle-aged stellar population of 2.5\,Gyr. 
For NGC~5092, the Lick index H$\delta_{\rm A}$, which is optimized for the stellar absorption from A stars (\citealp{1997ApJS..111..377W}), 
is $-1.7 \pm 0.6$\,{\angstrom} and the EW of the {\ha} emission line after correcting the stellar absorption is $1.1\pm 0.3$\,{\angstrom}.
By analyzing the host galaxy spectra of eight optical/UV TDEs and comparing their features to the SDSS main galaxy spectroscopic sample, \citet{2016ApJ...818L..21F}
find that six of eight TDE host galaxies, but only 2.3\% of local galaxies, have H$\delta_{\rm A} > 1.31$\,{\angstrom}, suggesting that TDEs may prefer 
the rare class of `E+A' galaxies, which are consistent with a post-starburst history. 
Compared with the TDE host galaxies discussed in \citet{2016ApJ...818L..21F}, NGC~5092 is redder and dominated by older stellar population, with H$\delta_{\rm A}=-1.7 \pm 0.6$\,{\angstrom}.
\citet{2017ApJ...835..176F} show that on the WHAN (\citealp{2010MNRAS.403.1036C}) diagram, at least
five of eight TDE host galaxies lie in the LINER-like region. This is not surprising owing to the possibility that 
interactions with the gas disk can increase the TDE rate by a factor of 10 (\citealp{2016MNRAS.460..240K}), 
and NGC~5092 is also located around the LINER-like region in the WHAN diagram.

\section{Summary}

We present and analyze the multiwavelength photometric and spectroscopic observations of the TDE candidate XMMSL1~J1319+2259 discovered in 
the {\it XMM-Newton} Slew survey in 2005 July, which shows contemporaneous
flaring in both X-ray and optical/UV bands. The source faded by a factor of $\sim$\,240 in the X-ray $0.2-2$\,keV band, 
about 1 mag in the optical band when it was reobserved 8 yr later, and 3 mag in the NUV band 3 yr later. 
The X-ray spectrum of the flare is soft and can be modeled with a blackbody at $kT \sim 60~\rm eV$ with an overall luminosity of 
$\sim$\,$1.5 \times 10^{43}$\,erg\,s$^{-1}$ ($L_{{\rm 0.01-100 {\rm keV}}}/L_{{\rm Edd}}\sim\,2.4 \times 10^{-3}$). 
In the optical--UV band, the emission kept rising from 2005 March to 2005 June and can 
be modeled with a blackbody at $T\sim (1-2) \times 10^4$\,K, with a luminosity of 
$\sim$\,$(2-6) \times 10^{43}$\,erg\,s$^{-1}$ ($L_{100-10000\,{\angstrom}}/L_{{\rm Edd}}\sim\,(3-9) \times 10^{-3}$), 
suggestive of an emission region of $\sim$\,100\,$R_{\rm g}$ in radius assuming a spherical geometry. 
Since the XMMSL observation in 2005 July is in the decline stage of the flare, the Eddington ratio during the UV observation in 2005 May 
should be $\gtrsim 0.01$. There was also a delayed continuous decline in the {\it WISE} infrared fluxes, from 5 to 13\,yr after the disruption, 
which can be described with a simplified infrared echo model with a dust temperature of 1000\,K. The host galaxy NGC~5092 is an 
early-type galaxy with a dominant stellar population of 2.5\,Gyr, older than that of the favorable 'E+A' type as suggested for TDE 
host galaxies (\citealp{2016ApJ...818L..21F}). The mass of the central BH is estimated to be $\sim$\,$5 \times 10^{7} \rm M_{\odot}$ from the $M_{\rm BH}$--$\sigma^*$ relationship. 
Analysis of the optical spectrum taken 3 yr after the flare reveals faint emission lines, which locate the galaxy at the borderline between AGN, starburst, 
and LINER in classification diagrams. While this does not allow unambiguous optical classification at present, we also cannot rule out the possibility 
that the emission lines are actually not  permanent, but were excited by the flare itself. We attribute the flare in the X-ray and optical/UV bands 
to a tidal disruption process, given the large-amplitude variabilities in multiple bands in the galactic nucleus and the ultra-soft outburst X-ray spectrum. 
XMMSL1~J1319+2259 is among the several TDEs that are detected to show bright flaring in both UV/optical and X-ray bands. 
Although containing few data points, TDEs discovered from archival data still provide important implications for statistical study, such as the event rate, 
the luminosity function, and the unification TDE model, as well as their host galaxy properties.
To discover larger samples of TDE, especially in real time, future high-cadence time-domain surveys are needed, 
such as the Large Synoptic Survey Telescope in optical (LSST; \citealp{2008SerAJ.176....1I}) and the {\it Einstein Probe} 
mission in the soft X-ray band (EP; \citealp{2016SSRv..202..235Y}, \citealp{2016IAUS..312...68Y}).

\section*{acknowledgements}
This work is supported by the National Natural Science Foundation of China (grant NOs. 11803047, 11833007, 11873054), 
the gravitational wave pilot B (grant NO. XDB23040100), and the Strategic Pioneer Program on Space Science, 
Chinese Academy of Sciences, grant NOs. XDA15310300, XDA15052100, and we acknowledge the support from the 
Faculty of the European Space Astronomy Centre (ESAC). Tinggui Wang is warmly thanked for his helpful suggestions and comments.
This work is mainly based on observations obtained with {\it XMM-Newton}, an ESA science mission with instruments 
and contributions directly funded by ESA Member States and NASA. This work also makes use of data products from ROSAT, 
{\it Swift}, {\it GALEX}, SDSS, {\it WISE}, and 2MASS. We use the ROSAT Data Archive of the Max-Planck-Institue f\"ur 
extraterrestrische Physik (MPE) at Garching, Germany. {\it GALEX} is operated for NASA by the California Institute 
of Technology under NASA contract NAS5-98034. {\it WISE} is a joint project of the University of California, Los Angeles, 
and the Jet Propulsion Laboratory/California Institute of Technology, funded by the National Aeronautics and Space Administration. 
2MASS is a joint project of the University of Massachusetts and the Infrared Processing and Analysis Center/California Institute of Technology, 
funded by the National Aeronautics and Space Administration and the National Science Foundation.
We acknowledge the use of public data from the {\it Swift} and SDSS data archive. 
\software{\textsc{sas} (v16.0.0; \citealp{2004ASPC..314..759G}), \textsc{xspec} (version 12.9.1; \citealp{1996ASPC..101...17A}), 
\textsc{gPhoton} (\citealp{2016ApJ...833..292M}), \textsc{magphys} (\citealp{2008MNRAS.388.1595D}), \textsc{galfit} (\citealp{2002AJ....124..266P})}

\appendix
\section{Reduction and Analysis of the multiwavelength data}
\label{sec:appendix}
In this appendix, we present detailed accounts of the data reduction and analysis in the X-ray (Appendix \ref{sec:app.x-ray}), optical (Appendix \ref{sec:app.optical}),
UV (Appendix \ref{sec:app.galex}) and infrared (Appendix \ref{sec:app.wise}). The details of the spectral analysis are described in Appendix \ref{sec.app:spectra}.

\subsection{X-Ray data}
\label{sec:app.x-ray}

We perform the {\it XMM-Newton} data reduction following 
the standard procedures with the software {\it XMM-Newton} Science Analysis System (\textsc{sas} version 16.0.0). 
XMMSL1~J1319+2259 was discovered in slew 9102500005 from XMM revolution 1025. 
With the slew data files (SDFs) downloaded from the {\it XMM-Newton} science archive\footnote{\url{http://nxsa.esac.esa.int/nxsa-web}},
the calibrated photon event files for pn are produced with \textsc{sas} task \texttt{eslewchain}. Since the source did not suffer 
from pileup problems in this observation, the pn spectrum was extracted with
patterns $0-4$. The source spectrum has been extracted from a circle of radius $ 30^{''}$ 
and a region of the same size was used for the background.
The technique described in \citet{2008A&A...482L...1R} was used to create the ancillary file, 
and the canned response file $eps\_ff20\_sdY6\_v6.9.rmf$ was used 
for the redistribution matrix.

The {\it XMM-Newton} pointed observation was carried out on 2013 December 21, 
with all the three {\it XMM-Newton} EPIC cameras operating in the Full Frame (FF) mode
and the thick filter in place. 
With the observation data files (ODFs),
the calibration files and calibrated photon event files for pn and MOS are produced with the 
tasks \texttt{epchain} and \texttt{emchain}. 
Source photons are extracted from a circular region with a $ 30^{''}$ radius centered on the object,
and background counts from a $ 60^{''}$ circular source-free region on the same CCD chip.
High background periods (count rates greater than 0.4/0.35 counts s$^{-1}$ for the pn/MOS camera)
caused by high-energy protons in orbit are filtered by creating good time intervals.
The pn spectrum is extracted from X-ray events with patterns from 0 to 4 
and the MOS spectra from 0 to 12. Ancillary files and response matrices are generated using 
the \texttt{arfgen} and \texttt{rmfgen} \textsc{sas} tasks, respectively. 

\subsection{Optical data}
\label{sec:app.optical}
\subsubsection{SDSS optical data}
\label{sec:app.sdss}
In the two SDSS observations taken on 2005 March 10 and 2005 April 2, the psField 
files\footnote{File with preliminary photometric calibration,
as well as final PSF fit, for a single field in an imaging run, which provides the 
2D reconstruction of the PSF.} and the drC images\footnote{The corrected image frame, having been bias subtracted, flat-fielded, 
purged of bright stars, and header updated with the latest calibrations.} are retrieved from the SDSS Data Archive Server\footnote{\url{das.sdss.org/www/html/das2.html}}. 
All the SDSS drC images have a `soft bias' of 1000 data numbers (DNs) added, so they can be stored as unsigned integers, as described
in the SDSS image processing pipeline (\citealp{2001ASPC..238..269L}).
With the reduced drC images, the aperture fluxes of the galaxy are measured using 
Source Extractor (SExtractor, version 2.25.0; \citealp{1996A&AS..117..393B}) within the aperture radii 
listed in the last column in Table \ref{tab:photometry}, which are chosen in such a way that the surface brightness 
beyond the radius does not change with the radial distance to the galactic center. 
The image subtraction process is as follows:
taking the $u$-band images as an example,
the PSF of the observed images is accessed by fitting the surface brightness of the unsaturated stars in the field
of view with a two-Gaussian model, giving an FWHM (full phrase) of 1$^{''}_{.}$47 and 1$^{''}_{.}$58 for the 2005-03 and 2005-04 images,respectively.
The images in the 2005-03 and 2005-04 observations are convolved with Gaussian profiles of FWHM=4.45\arcsec and 
4.43\arcsec, respectively, to have matching PSFs, resulting in smoothed images with nearly identical PSFs of 4.59\arcsec in FWHM.
The 2005-03 images are then scaled to the flux level of the 2005-04 images according to the counts measured of the nearby stars.
The residual images are generated by subtracting the rescaled 2005-03 images from the 2005-04 images.
The convolved images and the residual image after subtraction are shown in Figure \ref{fig:sdss_diff}.
There is a clear excess in the core of the galaxy in the 2005-04 image compared to that in 2005-03,
indicating brightening of the optical light, and thus a nonstellar process in the nuclear region. 

To quantify this optical flaring emission, we decompose the nuclear emission from the host galaxy for the drC images 
using the 2D imaging analysis tool \textsc{galfit} (\citealp{2002AJ....124..266P}).
The exact information about the PSF for each image is 
written out to the relevant psField files. We use the standalone code, read\_PSF, from the
READATLASIMAGES-V5\_4\_11 program,\footnote{\url{http://www.astro.princeton.edu/~rh1/readAtlasImages}} to reconstruct the PSF 
at the position of the galaxy NGC~5092 in the frame. The obtained PSF image has a standard size of 51 pixels on each side, and 
the background level is set to the standard soft bias of 1000 DN. For each drC image, a cutout of 200 pixels on each side centered on the core of NGC 5092
is used to ensure that enough pixels are
retained to properly determine the background. The 1000 DNs soft bias
is removed from the cutouts and PSF images prior to fitting.

The background is calculated by selecting a source-free region and is always included. 
The host galaxy is modeled with an exponential disk and a S$\acute{\mathrm{e}}$rsic bulge (the index fixed at 4). 
A point-like source, modeled by the PSF of the images, is added to 
account for any nonstellar nuclear emission.
For the 2005-03 observation, the stellar components (exponential disk and S$\acute{\mathrm{e}}$rsic)
are fixed to those fitted from the 2005-04 observation in all bands.
Such a model gives acceptable fits to all the images by visually examining the residual images. 
The estimation of the overall uncertainties of the fitted parameters is nontrivial (only statistical errors are given by \textsc{galfit}).
As a rough estimation, we compare the total fluxes of the model to the data fluxes for each of the images and find
differences  $< 30$~\%. We then simply take 30 per cent as the flux uncertainties of each model component in this work.
The offset of the fitted PSF and the nucleus is
subpixel (0$^{''}_{.}$396per pixel), or less than 150 pc. 

\subsubsection{XMM-{\it OM} and {\it Swift} UVOT data}
\label{sec:app.xmm}
With the ODF downloaded from the {\it XMM-Newton} science archive, we extract the background-subtracted photometric data using the \texttt{omichain}
processing pipeline with the default parameter setting as recommended by the \textsc{sas} threads.
The source partly overlaps the central ring of emission caused by the scattered background light and is larger than 
the upper limit of the extraction region, a circle with a radius of 24 pixels (corresponding to 11.4\arcsec), 
allowed by the interactive photometry task \texttt{omsource}.
So we define our own regions using the DS9 tool and perform aperture photometry on the
observing science window (OSW) IMAGE files obtained after running the \texttt{omichain} for each exposure.
The source is extracted in a circle of 20\arcsec radius, and 
the background is estimated in a circle with the same size in a source-free region.
Coincidence-loss, a phenomenon that occurs when the source count rate is high and the detector response will 
be nonlinear (\citealp{2000MNRAS.312...83F}), is corrected with the correction factor taken from the relevant 
exposure SWSRLI files. To convert the OM count rates into magnitudes and fluxes, the AB magnitude system for OM
implemented by the OM team is used\footnote{\textsc{sas} watch out page: \url{https://www.cosmos.esa.int/web/xmm-newton/sas-watchout-uvflux}.}.

For the UVOT observations, we follow the UVOT reduction threads\footnote{\url{https://www.swift.ac.uk/analysis/uvot/index.php}} 
for the data reduction work, with the 'Level 2' UVOT \textsc{FITS} files for each of the filters retrieved from the 
Space Science Data Center\footnote{\url{http://www.ssdc.asi.it/mmia/index.php?mission=swiftmastr}} used for aperture photometry.
Source counts are extracted from regions with a radius of 30\arcsec and  20\arcsec for U and UVW1 images, and a sky region of 
40\arcsec radius is used to estimate and subtract the sky background using the UVOT software task \texttt{uvotsource}.
To convert the count rates into magnitudes and fluxes, the most recent UVOT calibrations (\citealp{2008MNRAS.383..627P}; \citealp{2010MNRAS.406.1687B}) are used.

\subsection{{\it GALEX} UV data}
\label{sec:app.galex}

There were three detections with {\it GALEX} in NUV on 2005 May 28, 2005 June 11, and 2007 May 6 and one detection in FUV on 2007 May 6.
To generate calibrated light curves and images for the FUV and NUV observations, gPhoton (\citealp {2016ApJ...833..292M}) is used, 
which is a suite of publicly available {\it GALEX} calibration/reduction pipeline software tools written in Python.

We perform image subtraction on the 
calibrated NUV intensity images, generated with gPhoton image creation tool gMap with 0.5 degree angular extent
centered on the galaxy NGC 5092, between the 2005 May and 2007 May observations.  
The differential image exhibits a prominent excess---thus a flare in the NUV band---at the nucleus. We measure
the position of this excess emission using the SExtractor as R.A.=$13^{\rm h}19^{\rm m}51^s.5$, decl.=$23^{\circ}00^{'}00^{''}$.0, 
which is consistent with that of the galactic nucleus
(R.A.=$13^{\rm h}19^{\rm m}51^{\rm s}.5$, decl.=$22^{\circ}59^{'}59^{''}.6$)  
derived from $r$-band SDSS 2D image decomposition with \textsc{galfit} as described in Appendix \ref{sec:app.sdss}, within the positional uncertainty 
of 0.5$^{''}$ (\citealp{2007ApJS..173..682M}).

To quantify the amplitude of the variability, we measure the aperture photometry of the entire galaxy with gPhoton tool gAperture.
The photometry aperture is a circle of 20\arcsec radius, and the background is selected from an annularregion  with inner and outer radii of 25\arcsec and 30\arcsec, respectively.

\subsection{Infrared data}
\label{sec:app.wise}
The {\it WISE} has performed a full-sky imaging survey in four bands, $W1$ (3.4\,$\mu$m), $W2$ (4.6\,$\mu$m), $W3$ (12\,$\mu$m) 
and $W4$ (22\,$\mu$m), from 2010 February to August. {\it WISE} was in hibernation in 2011 February owing to the depletion of the 
solid hydrogen cryogen used to cool the $W3$ and $W4$ instrumentation, and was reactivated and
renamed NEOWISE-R since 2013 October, using only the $W1$ and $W2$.
We download the $W1$ and $W2$ data from 
the ALLWISE (\citealp{2010AJ....140.1868W}) and NEOWISE (\citealp{2014ApJ...792...30M}) multiepoch photometry data release 
by IRSA\footnote{\url{https://irsa.ipac.caltech.edu/frontpage/}}. 
The best-quality
single-frame images are used, by selecting only detections with good-quality frame  ($qi\_fact > 1$), low radiation hints ($saa\_sep > 5$),
not in the moon masking area ($moon\_masked = 0$), without contamination and confusion ($cc\_flag = 0$). 
The data mainly spread in 12 epochs at intervals of about 6 months, with each epoch containing about 12 single exposure 
observations performed within 1--2 days. To examine the long-term variability in IR, we smooth the light curves 
by measuring the median magnitudes and take the standard deviations as the magnitude uncertainties in each epoch. 
The magnitudes are then converted to the source flux density using the equations and parameters on the {\it WISE} 
data processing website\footnote {\url{http://wise2.ipac.caltecg.edu/docs/release/allsky/expsup/sec4\_4h.html}}.

NGC~5092 was also covered in 2MASS (\citealp{2006AJ....131.1163S}),
and the Atlas images in each filter, used for aperture photometry,
are accessible via the 2MASS Images Services,\footnote{\url{https://irsa.ipac.caltech.edu/applications/2MASS/IM/}} 
which are administered by the Infrared Science Archive (IRSA)\footnote{\url{https://irsa.ipac.caltech.edu/frontpage/}}. 
To convert the calibrated magnitudes into
fluxes, the fluxes for zero-magnitude conversion values from \citet{2003AJ....126.1090C} are used. 

\subsection{Optical spectral analysis}
\label{sec.app:spectra}
\subsubsection{Starlight modeling}
\label{sec:app.starlight} 
A series of procedures, which have been developed and applied in our 
previous work (\citealp{2006AJ....131..790L}; \citealp{2006ApJS..166..128Z}; \citealp{2018ApJS..235...40L}),
are used to deal with starlight and continuum fitting in this work.
The starlight is modeled using a combination of six 
synthesized galaxy templates derived from the spectral template 
library of simple stellar populations (SSPs; \citealp{2003MNRAS.344.1000B}) 
using the Ensemble Learning Independent Component Analysis 
algorithm (see \citealp{2006AJ....131..790L} for details\footnote{These 
galaxy templates have been applied in a series of spectral 
analyses in our previous work (e.g. \citet{2006ApJS..166..128Z}, \citet{2012ApJ...755..167D}, \citet{2018ApJS..235...40L}, 
and demonstrated to be robust and effective in dealing with 
large dataset of spectra.}). These templates have included 
most stellar features in SSPs and hence can avoid underlying 
overfit in most cases. These templates are broadened by 
convolving with a Gaussian of an alterable width to match 
the stellar velocity dispersion of the host galaxy. In addition,
the starlight model is slightly shifted in wavelength in adaptive steps 
to correct the effect introduced by the possible uncertainty of 
the redshift provided by the SDSS pipeline, which can help 
to improve the fit and measurement of the host velocity 
dispersion. During the continuum fitting, two kinds of regions are masked out: (1) bad pixels as flagged by 
the SDSS pipeline, and (2) wavelength range that may be seriously affected by prominent emission lines. A composite SDSS quasar spectrum introduced 
by \citet{2001AJ....122..549V} is initially adopted to determine the second mask regions. The starlight can be well modeled with a stellar population of 2.5~Gyr, and
the fitting result is shown in Figure \ref{fig:host optical spectrum}.
\subsubsection{Emission-line measurements}
\label{sec:app.emission}
After the subtraction of the stellar continuum, a residual pure emission-line  spectrum is obtained, which is fitted as follows. Each of 
the narrow lines, including {\ha}, {\hb}, {\oiii}, and 
{\nii}, are fitted using a Gaussian profile; the profiles and 
redshifts of these narrow lines are tied to those of narrow {\oiii}, 
and the flux ratios of {\oiii} and {\nii} doublets are fixed to their theoretical values, respectively. 
We note that the emission-line-fitting results are vulnerable to continuum subtraction. Thus, in order to check the significance of these emission
lines, we try to estimate the uncertainties as described below. We find a series of fake emisson-line features from the residual spectrum where there are no
prominent emission lines in theory. These so-called emission lines are fitted with one Gaussian, 
whose widths are fixed to that of {\oiii}. The standard deviation of the derived fluxes of these fake 
lines is then taken as $1\sigma$ uncertainty of the detected emission lines. As a result, {\oiii}$\lambda5007$ and 
{\ha} are significantly detected with $S/N > 3$, while {\hb} is marginally detected with $S/N \approx 2.4$.

\subsubsection{Upper Limits estimation}
\label{sec:app.upperlimit}
\begin{figure*}
 \centering
 \includegraphics[width=0.6\textwidth]{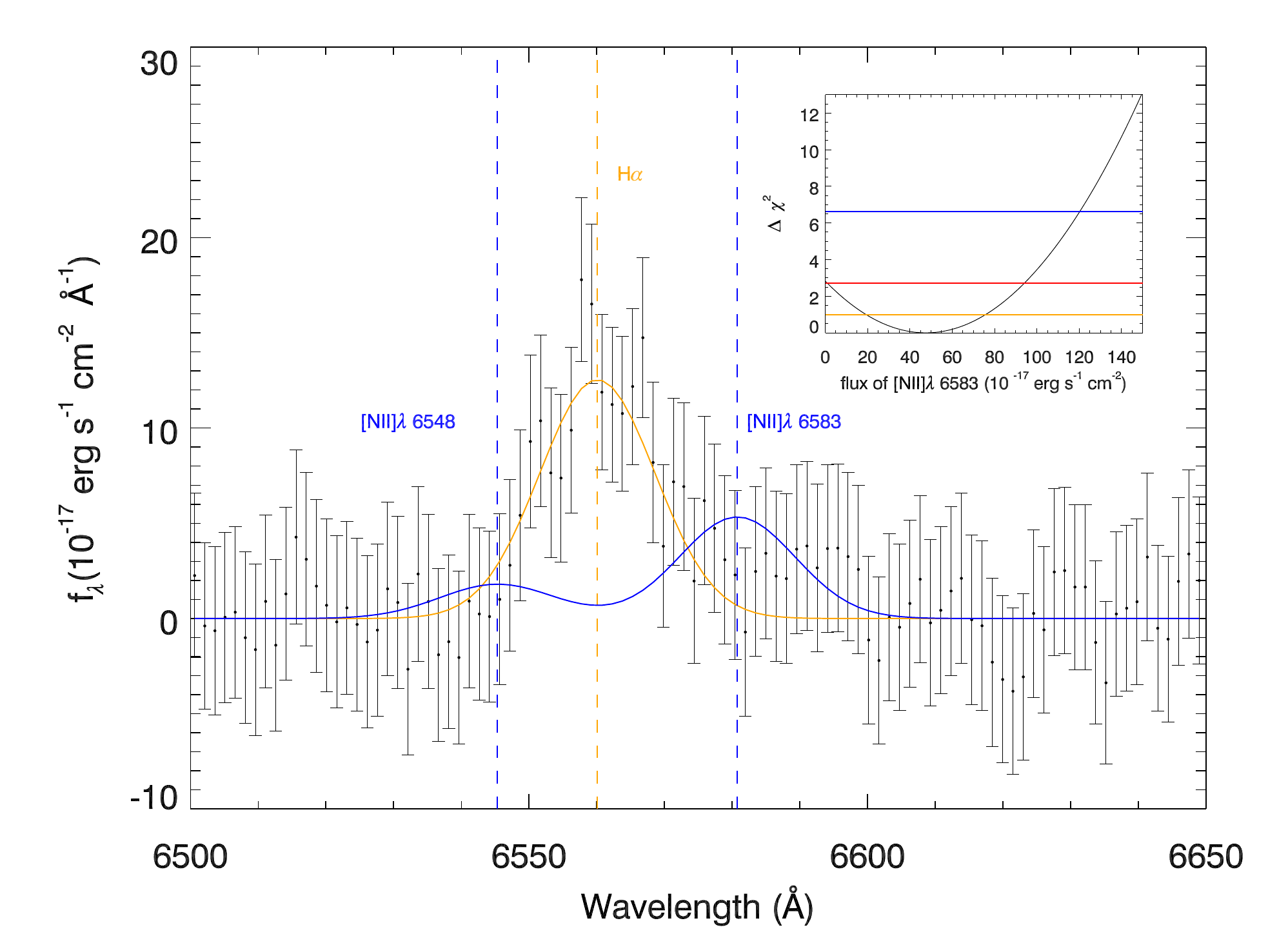}
 \caption{ The optical starlight-model-subtracted spectrum in the range of $6500-6650$\,\AA, with the best-fit 
{\ha} (orange) and the 90\% upper limits of the {\nii} emission lines (blue). The inset plot shows the 
fitted $\Delta \chi^2$ versus the flux of {\nii} $\lambda 6583$, with the horizontal lines indicating the 68\% (orange), 90\% (red), 
and 99\% (blue) confidence levels, respectively.}
 \label{fig: upper limit}
\end{figure*}

For lines of interest without robust detections, upper limits on their fluxes are derived by using the $\chi^2$ 
fitting technique, following the method used in \citet{2015AJ....149...75L} (see their Section 2.1).
Here we take the measurement of the flux upper limit of {\nii} as an example.
The emission-line spectrum segment in a wavelength range covering the {\nii}+{\ha} region (6500--6650\AA) is used only. 
First, the line complex is modeled with a Gaussian for {\ha} and two Gaussians for the {\nii} doublets. 
The spectral segment is fitted with this model by using the $\chi^2$ fitting procedure, resulting in the minimum $\chi^2_{min}$ 
and a set of best-fit parameters. We then vary the flux of {\nii} over a wide range with an increment and fix it in a series of fits, 
resulting in a series of gradually increasing $\chi^2$ values. The 90\% confidence range of the flux is determined by 
finding the corresponding $\Delta \chi^2 = \chi^2-\chi^2_{min}$ = 2.706 (for one parameter of interest), 
which is 9.4$\times$10$^{-16}$\,erg\,s$^{-1}$\,cm$^{-2}$ for {\nii}$\lambda6583$.
Figure \ref{fig: upper limit} shows the starlight-model-subtracted spectrum in the wavelength 
range of $6500-6650$\,\AA, with the best-fit {\ha} and 90\% upper limits of the {\nii} emission lines (blue). 
The inset plot shows the $\Delta \chi^2$ versus the flux of {\nii} $\lambda 6583$ and the orange, red and green horizontal lines
indicate the 68\%,90\%, and 99\% confidence levels, with $\Delta \chi^2~=1.0, 2.706~\rm and ~6.63$~respectively.

\newpage
\bibliographystyle{aasjournal}
\bibliography{ngc5092}

\end{document}